\documentclass[aps,prb,twocolumn,showpacs,10pt,floatfix,superscriptaddress,floatfix]{revtex4-2}
\usepackage{epsf}
\usepackage{bm}
\usepackage{hyperref}
\usepackage{amsfonts}
\usepackage{amssymb}
\usepackage{amsmath,mathtools}
\usepackage{array}
\usepackage{enumerate,dsfont}
\usepackage{dcolumn,multirow}
\usepackage[utf8]{inputenc}
\usepackage{latexsym}
\usepackage{xcolor}

\newcommand{\vk}{\vec k}

\newcommand{\ZZ}{\mathbb{Z}}

\usepackage{ bbold }

\renewcommand{\vec}[1]{\mathbf{#1}}

\begin{document}

\title{Higher-order topological semimetals and nodal superconductors with an order-two crystalline symmetry}

\author{Sophia Simon}
\affiliation{Dahlem Center for Complex Quantum Systems and Physics Department, Freie Universit\"at Berlin, Arnimallee 14, 14195 Berlin, Germany}
\affiliation{Department of Physics, University of Toronto, Toronto, Ontario M5S1A7, Canada}
\author{Max Geier}
\affiliation{Dahlem Center for Complex Quantum Systems and Physics Department, Freie Universit\"at Berlin, Arnimallee 14, 14195 Berlin, Germany}
\affiliation{Center for Quantum Devices, Niels Bohr Institute, University of Copenhagen, DK-2100 Copenhagen, Denmark}
\author{Piet W. Brouwer}
\affiliation{Dahlem Center for Complex Quantum Systems and Physics Department, Freie Universit\"at Berlin, Arnimallee 14, 14195 Berlin, Germany}
\date{\today}

\begin{abstract}
Using a systematic relation between topological gapless phases in three dimensions and topological gapped phases in two dimensions, we identify four types of higher-order topological semimetals or nodal superconductors (HOTS), hosting (i) flat zero-energy ``Fermi arcs'' at crystal hinges, (ii) flat zero-energy hinge arcs coexisting with surface Dirac cones, (iii) chiral or helical hinge modes, or (iv) flat zero-energy hinge arcs connecting nodes only at finite momentum. Bulk-boundary correspondence relates the hinge states to the bulk topology protecting the nodal point or loop. We classify all HOTS for all tenfold-way classes with an order-two crystalline (anti-)symmetry, such as mirror, twofold rotation, or inversion.
\end{abstract}
\maketitle

\section{Introduction}
Topological semimetals have a band structure with a touching of valence and conduction bands that is robust to small parameter changes that preserve the crystalline symmetries \cite{volovik2009,vafek2014,burkov2016,fang2016,yan2017,armitage2018,burkov2018, gao2019r}.
The prime example is a Weyl semimetal, which has twofold degenerate band crossing points around which the band structure is effectively described in terms of massless Weyl fermions \cite{ haldane2004, murakami2007, wan2011, yang2011, burkov2011a, xu2011, burkov2011b, young2012, wang2012, wang2013, hasan2017r}. Similarly, topological nodal superconductors have protected gapless modes in the excitation spectrum of the Bogoliubov quasiparticles \cite{volovik2009,schnyder2015,volovik2013,sato2017}. 
A topological classification of nodal points or lines, indicating which type of node may be protected for which combination of symmetries, has been obtained for the ``tenfold-way'' symmetry classes \cite{altland1997}. These are defined by the presence or absence of time-reversal and spin-rotation symmetry, as well as the particle-hole symmetry associated with the mean-field description of superconductors \cite{schnyder2008}. The results have been subsequently extended to incorporate crystalline symmetries \cite{shiozaki2014, chiu2014, shiozaki2017, shiozaki2018atiyah, cornfeld2019, shiozaki2019classification, cornfeld2021, ono2021nodal}. 

Topological nodal semimetals and superconductors may have gapless surface excitations, with a surface band structure that connects to the projections of the nodes of the bulk band structure onto the surface Brillouin zone. These surface excitations are anomalous, in the sense that they cannot be removed by a local, symmetry-allowed perturbation \cite{matsuura2013}.
For example, a Weyl semimetal has a ``Fermi-arc'' of surface states,
whereas nodal-loop semimetals have ``drumhead'' surface states, an approximately flat ``Fermi disk'' of surface states inside the projection of the nodal loop on the surface \cite{burkov2011b,fang2015nodal, rui2015, huang2016, schoop2016, chan2016, fang2016, yu2017r, yang2018}.

There is an insightful theoretical argument that links the protected surface states and nodal points or lines in gapless band structures to properties of {\it gapped} topological band structures in lower dimensions. For example, the Hamiltonian $H(k_x,k_y,k_z)$ of a Weyl semimetal can be considered as a $k_z$-dependent ``one-parameter family'' of two-dimensional Hamiltonians $H_{k_z}(k_x,k_y)$.
A topological phase transition of $H_{k_z}(k_x,k_y)$ as a function of $k_z$ --- a ``band inversion'' --- then corresponds to a protected gap closing in three dimensions \cite{hughes2011,xu2011}, whereas
the chiral edge states in the topological phase of $H_{k_z}(k_x,k_y)$ naturally map to the Fermi-arc surface states \cite{wan2011,chiu2016, hasan2017r}. In the same way, a gapless band structure with a nodal loop can be obtained by considering $H(k_x,k_y,k_z)$ as a ``two-parameter family'' of one-dimensional Hamiltonians \cite{horava2005, fang2016, yang2018}. The flat disk of ``drumhead'' surface states is then associated with the protected end-states of topological band structures in one dimension \cite{matsuura2013}.

With crystalline symmetries, anomalous boundary states may also exist at edges or corners of a crystal, instead of being extended across the entire surface \cite{schindler2018hoti,benalcazar2017b,langbehn2017,song2017,fang2019,geier2018,khalaf2018inversion, khalaf2018symmetry,trifunovic2019,okuma2019,ahn2019,roberts2019,kooi2019,rasmussen2018}. Band structures that impose such boundary signatures are said to possess ``higher-order'' topology \cite{schindler2018hoti,trifunovic2020}.
For a large class of crystalline symmetries, classification information of gapped topological band structures that includes the order of the topological phase is available \cite{geier2018,khalaf2018symmetry,khalaf2018inversion,trifunovic2019,okuma2019,ono2020,geier2020,geier2021, ono2021Z2}.

\begin{figure*}
\includegraphics[width=2\columnwidth]{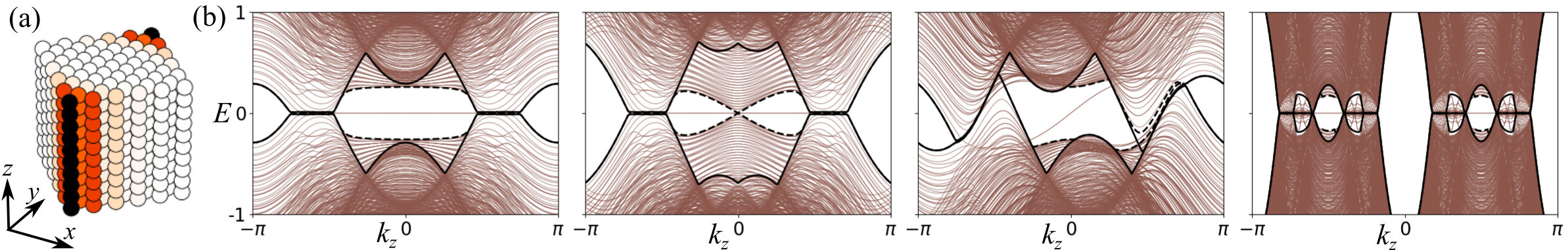}
\caption{\label{fig:bandstructures}
  (a) Rhombic pillar geometry and local density of states at energy $E = 0$ of model \eqref{eq:H_AIII_My+}. (b) Quasi one-dimensional band structures in the rhombic pillar geometry for the models of Eqs. \eqref{eq:H_AIII_My+}, \eqref{eq:H_DIII_M--}, \eqref{eq:H_D_Rz-}, and \eqref{eq:H_DIII_Rz-+} (from left to right).
The full and dashed black curves indicate the bulk and surface excitation gaps, respectively. In models \eqref{eq:H_AIII_My+}, \eqref{eq:H_DIII_M--}, and \eqref{eq:H_D_Rz-}, we set $t=1$ and $k_0 = 0.6\pi$. In model \eqref{eq:H_DIII_Rz-+} we used $m=0$, $t=1$, $t'=2$. } 
\end{figure*}

Soon after the discovery of gapped higher-order topological band structures, it was realized that topological gapless phases may also have higher-order anomalous boundary states \cite{ezawa2018,lin2018semimetal,wang2019,wang2020b,rui2021,zhao2020,wieder2020,wang2020a,ghorashi2020,
zhang2020,calugaru2019,roy2019,ghorashi2019,wei2021,ni2021,luo2021,lange2021,zhaoPRL2021}. In particular, a second-order topological semimetal has anomalous states on crystal hinges, extending between the projections of the nodal points of the bulk band structure. No third-order topological semimetals can exist, because a putative anomalous corner state can always hybridize with the continuum of bulk states.

In this article, we consider the question, for which combinations of tenfold-way and crystalline symmetries second-order topological gapless phases may exist. Hereto, we note that, if we again view the Hamiltonian $H(k_x,k_y,k_z)$ as a one-parameter family of two-dimensional Hamiltonians $H_{k_z}(k_x,k_y)$, a second-order topological nodal semimetal or superconductor arises if, as a function of $k_z$, $H_{k_z}(k_x,k_y)$ goes through a topological phase transition involving a gapped second-order phase atomic-limit phase \cite{lin2018semimetal,wieder2020}
 \footnote{ We do not consider ``higher-order Weyl semimetals'', for which, as a function of $k_z$, there is a topological phase transition between a first-order and a second-order phase \cite{ghorashi2020,wang2020b,luo2021}. Such topological band structures may be viewed as the direct sum of a standard Weyl semimetal and a gapped second-order topological band structure.}.
 Hence, to determine which combinations of tenfold-way and crystalline symmetries allow for second-order nodal phases, it suffices to inspect topological phase transitions of the two-dimensional Hamiltonians $H_{k_z}(k_x,k_y)$. We here provide such a systematic classification for three crystalline symmetries: mirror, twofold rotation, or inversion symmetry. These are the symmetries for which a full order-resolved classification of gapped topological band structures exists \cite{geier2018,khalaf2018symmetry,khalaf2018inversion,trifunovic2019}.

Specifically, we consider a pillar geometry, translation invariant in the $z$ direction and finite, with symmetry-compatible lattice termination, in the $xy$ plane, see Fig.\ \ref{fig:bandstructures}(a). We set the unit cell size along $z$ to $1$, so that $-\pi < k_z \le \pi$. We view the Hamiltonian $H(k_x,k_y,k_z)$ of such a crystal as a $k_z$-dependent family $H_{k_z}(k_x,k_y)$, where each Hamiltonian $H_{k_z}(k_x,k_y)$ describes a two-dimensional crystal with symmetry-compatible boundaries. Throughout, we take $H_0(k_x,k_y)$ and $H_{\pi}(k_x,k_y)$ to be gapped, so that any nodal points or loops in the bulk band structure occur away from the high-symmetry planes $k_z=0$, $\pi$. For definiteness, we assume that $H_{\pi}(k_x,k_y)$ is topologically trivial, which is an assumption that can be made without loss of generality for a classification of features directly associated with the nodes of the bulk spectrum. In this geometry, anomalous second-order boundary states (if any) exist between projections of bulk nodes on both sides of $k_z = 0$ or between projections of two bulk nodes on the same side of $k_z=0$.

The tenfold-way and crystalline symmetries imply constraints for the family of two-dimensional Hamiltonians $H_{k_z}(k_x,k_y)$.
As symmetry operations may change the sign of $k_z$, in general the symmetry constraints at the high-symmetry planes $k_z=0$, $\pi$ differ from those for generic $0 < |k_z| < \pi$. Because of the different symmetries involved, different types of anomalous boundary states are possible around $k_z = 0$ and for generic $0 < |k_z| < \pi$. Considering all combinations of tenfold-way and order-two crystalline symmetries, we find three scenarios for topological semimetals or nodal superconductors with second-order hinge states around $k_z=0$ and one scenario for second-order hinge states in a region at finite $0 < |k_z| < \pi$. We now discuss these four scenarios, together with examples of lattice models that realize them.

\section{Four types of second-order topological semimetals and nodal superconductors}

{\it (i) Flat zero-energy hinge arcs around $k_z = 0$.---} If $H_{k_z}(k_x,k_y)$ is in a second-order topological phase for $k_z = 0$ and remains so upon (continuously) going to finite $|k_z| > 0$, there are flat arcs of zero-energy hinge states running between the projections of nodal points or loops on both sides of $k_z = 0$ onto the edge Brillouin zone. (The hinge states are the zero-energy corner states of the two-dimensional Hamiltonians $H_{k_z}(k_x,k_y)$.) An example is a band structure with a chiral antisymmetry $\mathcal{C}$, corresponding to tenfold-way class AIII, and a mirror symmetry ${\cal M}^x$ that commutes with ${\cal C}$. This symmetry class may be realized in a time-reversal symmetric superconductor with $\text{U}(1)$ spin-rotation symmetry \cite{schnyder2008}. A concrete lattice model is the four-band model
\begin{equation}
 \begin{split}
 H(k_x,k_y,k_z) = & \, t \tau_1 \sigma_1 [m(k_z) + 2 - \cos k_x - \cos k_y] \\
 &\, \mbox{} + t \tau_1 \sigma_3 \sin k_x + t \tau_1 \sigma_2 \sin k_y,
\end{split}
\label{eq:H_AIII_My+}
\end{equation}
with Pauli matrices $\tau_j$ and $\sigma_j$, $j=0,1,2,3$. Here and below, $m(k_z)$ is defined as $m(k_z) = \cos k_0 - \cos k_z$.

The symmetry constraints are
\begin{align}
  H(k_x,k_y,k_z) =&\, - U_{\cal C} H(k_x,k_y,k_z) U_{\cal C}^{\dagger} \nonumber \\ = &\,
  U_{\cal M} H(-k_x,k_y,k_z) U_{\cal M}^{\dagger},
  \label{eq:symm_AIII_My+}
\end{align}
with $U_\mathcal{C} = \tau_3 \sigma_0$ and $U_{\mathcal{M}} = \tau_3 \sigma_3$. The Hamiltonian (\ref{eq:H_AIII_My+}) has double band crossings at $k_z = \pm k_0$, which are turned into nodal loops upon addition of a generic symmetry-preserving perturbation. 
Seen as a two-dimensional Hamiltonian, $H_{k_z}(k_x,k_y)$ is in a second-order phase with non-degenerate zero-energy states at mirror-symmetric corners for $-k_0 < k_z < k_0$ and in a trivial phase for $|k_z| > k_0$, see App.\ \ref{app:b} for details. Hence, for a pillar geometry the three-dimensional Hamiltonian $H(k_x,k_y,k_z)$ describes a nodal-loop semimetal with a flat arc of hinge states. Exact diagonalization, with symmetry-preserving terms added such that eventual trivial surface states are gapped out, confirms the appearance of a flat arc of zero-energy states on the two mirror-symmetric hinges of the pillar, see Figs.~\ref{fig:bandstructures}(a) and (b).
The symmetry-preserving perturbation used in the calculations of Fig.\ \ref{fig:bandstructures} has the form 
\begin{equation}
  \delta H(k_x,k_y,k_z) = b \tau_2 \sigma_2
  \label{eq:Hadd}
\end{equation}
 with $b = 0.4 t$. In addition to gapping out non-topological surface states, the addition of this term also lifts the accidental degeneracy of the nodal point at $k_z = \pm k_0$ and turns it into a nodal loop, see App.\ \ref{app:b}.

The presence of the antisymmetry ${\cal C}$ is essential for pinning the hinge states to zero energy. Other antisymmetries, such as the antiunitary particle-hole antisymmetry ${\cal P}$ associated with the Bogoliubov-de Gennes description of superconductors or combinations of ${\cal C}$ or ${\cal P}$ and crystalline symmetries can also stabilize hinge states at zero energy. Flat arcs of hinge states also appear in proposals for second-order semimetals in the literature \cite{ezawa2018,lin2018semimetal,wang2019,zhao2020,ghorashi2020,wieder2020,wang2020a,wang2020b,rui2021,wei2021,ni2021,zhaoPRL2021}, although the antisymmetry required for the topological protection is not always recognized explicitly \cite{ezawa2018,lin2018semimetal,wang2019,zhao2020,wang2020b,rui2021,ni2021,zhaoPRL2021} or is allowed to be weakly broken \cite{wieder2020,ghorashi2020,wei2021}, so that the hinge arc is no longer strictly pinned to zero energy. In that case, the hinge arc no longer has a strict topological protection and may be removed by a sufficiently strong local perturbation at the crystal hinge. 

The tenfold-way class CII with a twofold rotation symmetry ${\cal R}^z$ allows for a variant of this scenario, in which the zero-energy hinge arcs do not terminate at nodes of the bulk band structure, but at (topologically protected) Dirac nodes of the surface band structure. In this case, the presence of anomalous zero-energy hinge states at $k_z = 0$ as well as their absence at $k_z = \pi$ are protected by the topology of the bulk band structure, whereas the zero-energy hinge states at generic $0 < |k_z| < \pi$ are ``extrinsic'', {\em i.e.} unrelated to the bulk topology \cite{geier2018}. Such hinge states may disappear at a $k_z$-dependent closing of the surface gap. We refer to App.\ \ref{app:c} for details.

{\it (ii) Flat zero-energy hinge arcs around $k_z = 0$ and first-order surface states at $k_z = 0$.---} With a suitable combination of tenfold-way and crystalline symmetries, it is possible that $H_{k_z}(k_x,k_y)$ is in a first-order topological phase at $k_z = 0$, and in a second-order topological phase at finite $0 < |k_z| < k_0$. Then, the band structure is of a hybrid order, with a flat arc of zero-energy states at the hinge coexisting with first-order Dirac cone surface states on adjacent surfaces that are themselves not invariant under the crystalline symmetry.
An example is a spinful odd-parity time-reversal symmetric superconductor (tenfold-way class DIII) with 
mirror symmetry ${\cal M}$.
A concrete model is the four-band Hamiltonian
\begin{equation}
\begin{split}
 H(k_x,k_y,k_z) = & \, t \tau_3 \sigma_0 [m(k_z) + 2 - \cos k_x - \cos k_y] \\
 &\, \mbox{} + t \tau_2 \sigma_0 \sin k_x + t \tau_1 \sigma_3 \sin k_y,
\end{split}
\label{eq:H_DIII_M--}
\end{equation}
which has nodal points at $|k_z| = k_0$ and satisfies the symmetry constraints
\begin{align}
  \label{eq:DIIIsymmetry}
  H(k_x,k_y,k_z) =&\, - U_{\cal P} H(-k_x,-k_y,-k_z)^* U_{\cal P}^{\dagger} \nonumber \\ =&\,
  U_{\cal T} H(-k_x,-k_y,-k_z)^* U_{\cal T}^{\dagger} \nonumber \\ =&\,
  U_{\cal M} H(-k_x,k_y,k_z) U_{\cal M}^{\dagger},
\end{align}
where $U_\mathcal{T} = \tau_0 \sigma_2$, $U_\mathcal{P} = \tau_1 \sigma_0$, 
and $i U_\mathcal{M} = i \tau_3 \sigma_1$ represent time reversal, particle-hole conjugation, and mirror, respectively.
At generic $k_z$, $H_{k_z}(k_x,k_y)$ is invariant under the products ${\cal P} {\cal T}$ and ${\cal T} {\cal M}$ that leave $k_z$ invariant, but not under ${\cal P}$, ${\cal T}$, or ${\cal M}$ separately. Because the symmetries differ for $k_z = 0$ or $\pi$ and generic $0 < |k_z| < \pi$, $H_{k_z}(k_x,k_y)$ is in different topological classes at $k_z = 0$ and at $k_z \neq 0$: a first-order phase with helical Majorana edge states for $k_z = 0$, and a second-order phase with zero-energy corner states for $0 < |k_z| < k_0$. As shown in Fig.\ \ref{fig:bandstructures}(b), for the three-dimensional Hamiltonian $H(k_x,k_y,k_z)$ this implies a combination of Dirac cone surface states at $k_z=0$ and a flat hinge arc ranging from $k_z = 0$ to the projection of the nodal loop around $|k_z| = k_0$ onto the hinge Brillouin zone.
To remove non-topological surface states and accidental degeneracies, the additional term 
\begin{equation}
  \delta H(k_x,k_y,k_z) = b \tau_0 \sigma_1 \sin k_z
  \label{eq:Hadd_DIII_M--}
\end{equation}
 with $b = 0.4 t$ was added to the model \eqref{eq:H_DIII_M--}.
 
{\it (iii) Chiral or helical hinge states around $k_z = 0$.---} With rotation or inversion symmetry, it is possible that $H_{0}(k_x,k_y)$ is in a second-order phase with zero-energy corner states, whereas $H_{k_z}(k_x,k_y)$ is in a nontrivial atomic limit phase at generic $k_z$. In this case, the bulk band crossing is protected by the transition between the nontrivial and trivial atomic limits as a function of $k_z$. Around $k_z = 0$, such a system hosts chiral or helical hinge mode which are pinned to zero energy at $k_z = 0$.

As an example, we consider a crystal with twofold rotation symmetry ${\cal R}^z$, which has co-propagating chiral hinge modes on rotation-related hinges. Co-propagating chiral hinge modes are ruled out in gapped topological phases, but with a gapless bulk, an equilibrium current carried by bulk states may compensate the equilibrium current carried by the chiral hinges. This scenario applies to a spinful even-rotation-parity time-reversal symmetry-breaking superconductor (tenfold-way class D), which has the Bogoliubov-de Gennes Hamiltonian
\begin{equation}
\begin{split}
   H(k_x,k_y,k_z) = & \, t \tau_3 \sigma_3 [m(k_z) + 2 - \cos k_x - \cos k_y] \\
   &\, \mbox{} + t \tau_1 \sigma_3 \sin k_x + t \tau_2 \sigma_3 \sin k_y,
\end{split}
\label{eq:H_D_Rz-}
\end{equation}
with the representations $U_\mathcal{P} = \tau_1 \sigma_0$ and $i U_\mathcal{R} = i \tau_3 \sigma_3$.
Since particle-hole conjugation changes the sign of $k_z$, for generic $k_z$, $H_{k_z}(k_x,k_y)$ is constrained by ${\cal R}_z$ only.
The spectrum, showing the linearly dispersing chiral hinge states around $k_z = 0$, is shown in Fig.\ \ref{fig:bandstructures}(b).
In the presence of time-reversal symmetry, this scenario leads to helical hinge states around $k_z=0$ instead of chiral hinge states, as discussed in App.\ \ref{app:c}.
To remove non-protected surface states, we added an additional term of the form 
\begin{equation}
  \delta H(k_x,k_y,k_z) = b_1 \sigma_2 \tau_2 + b_2 \sigma_1 \tau_1 \sin k_z
  \label{eq:Hadd_D_Rz-}
\end{equation}
 with $b_1 = b_2 = 0.4 t$ to the model Hamiltonian \eqref{eq:H_D_Rz-}. As in the previous cases, this additional term also lifts the accidental degeneracy of the nodal point at $k_z = \pm k_0$ and turns it into a nodal loop. We refer to App.\ \ref{app:b} for a more detailed discussion. 

{\it (iv) Flat zero-energy hinge arcs away from $k_z = 0$.---} If $H_{k_z}(k_x,k_y)$ can be in a second-order phase for generic $0 < |k_z| < \pi$, flat zero-energy hinge arcs between projections of bulk nodes on hinge Brillouin zones away from $k_z = 0$ and $k_z = \pi$ are possible irrespective of the topological phase of $H_{k_z}(k_x,k_y)$ at $k_z = 0$. Such flat arcs of zero-energy hinge states are the only possible second-order boundary features associated with a bulk gap closing if the topological classification of $H_{k_z}(k_x,k_y)$ at $k_z = 0$ does not allow for topological phases with anomalous boundary features or $H_{k_z}(k_x,k_y)$ immediately becomes trivial upon going from $k_z = 0$ to nonzero $k_z$.

An example is a spinful odd-rotation-parity time-reversal-symmetric superconductor (class DIII)
\begin{align}
  H(k_{x},k_{y},k_{z}) =&\, t \tau_{0}\sigma_{3}
  \sin k_z (m-1-\cos k_{x}-\cos k_{y}) \nonumber \\ 
  &\, \mbox{}+ t \tau_{0}\sigma_{1}\sin k_{x} + t \tau_{3}\sigma_{2}\sin k_{y} \nonumber \\
  &\, \mbox{}+ t'  \tau_3 \sigma_0 \cos^2 k_z,
\label{eq:H_DIII_Rz-+}
\end{align}
with the representations $U_\mathcal{T} = \tau_0 \sigma_2$, $U_\mathcal{P} = \tau_1 \sigma_0$, and $i U_{\mathcal{R}} = i \tau_0 \sigma_3$. The Hamiltonian (\ref{eq:H_DIII_Rz-+}) is topologically trivial at $k_z = 0$, $\pi$ and is in a second-order phase around $|k_z| = \pi/2$. It has a total of eight nodal loops. Four of these, which appear closest to $k_z = 0$ and $\pi$, are ``fragile'' and can be removed by the addition of trivial bands (see App.\ \ref{app:b}). The remaining four, which are closer to $k_z = \pm \pi/2$, are associated with the topological phase transitions between second-order and atomic-limit phases of $H_{k_z}(k_x,k_y)$. There is a flat zero-energy hinge arc connecting projections of the nodal loops around $k_z = \pm \pi/2$ onto the edge Brillouin zone. Figure \ref{fig:bandstructures}(b) shows the spectrum for this model for the pillar geometry. 
For the numerical calculations of Fig.\ \ref{fig:bandstructures}(b) we added the term 
\begin{equation}
  \delta H = b \tau_2 \sigma_0 \sin k_z
\end{equation}
with $b = 0.4 t$ to the model \eqref{eq:H_DIII_Rz-+} to gap out non-protected surface states.

Further illustrative examples for each type are presented in App.\ \ref{app:c}.

\begin{table}
\begin{tabular}{cccc|c|cccc}
Class & $\mathcal{T}^{2}$ & $\mathcal{P}^{2}$ & $\mathcal{C}^{2}$ & type & $\mathcal{M}^{x/y}$ & $\mathcal{R}^{x/y}$ & $\mathcal{R}^{z}$ & $\mathcal{I}$\tabularnewline
\hline 
$\text{AIII}$ & 0 & 0 & 1 & (i) & $\mathcal{M}_{+}^{x/y}$ & $\mathcal{T}^{+}\mathcal{R}_{+}^{x/y}$ & $\mathcal{R}_{-}^{z}$ & $\mathcal{T}^{+}\mathcal{I}_{-}$\tabularnewline
BDI & 1 & 1 & 1 & (i) & $\mathcal{M}_{++}^{x/y}$ & $\mathcal{R}_{++}^{x/y}$ & $\mathcal{R}_{+-}^{z}$ & $\mathcal{I}_{+-}$\tabularnewline
 &  &  &  & (iv) & $\mathcal{M}_{--}^{x/y}$ &  & $\mathcal{R}_{-+}^{z}$ & \tabularnewline
D & 0 & 1 & 0 & (i) & $\mathcal{T}^{+}\mathcal{M}_{+}^{x/y}$ & $\mathcal{R}_{+}^{x/y}$ &  & \tabularnewline
 &  &  &  & (iii) &  &  & $\mathcal{R}_{-}^{z}$ & $\mathcal{T}^{+} \mathcal{I}_{+}$\tabularnewline
 &  &  &  & (iv) & $\mathcal{T}^{-}\mathcal{M}_{-}^{x/y}$ &  &  & \tabularnewline
DIII & -1 & 1 & 1 & (i) & $\mathcal{M}_{--}^{x/y}$ &  &  & \tabularnewline
 &  &  &  & (ii) & $\mathcal{M}_{--}^{x/y}$ & $\mathcal{R}_{-+}^{x/y}$ & $\mathcal{R}_{+-}^{z}$ & $\mathcal{I}_{--}$\tabularnewline
 &  &  &  & (iii) &  &  & $\mathcal{R}_{+-}^{z}$ & \tabularnewline
 &  &  &  & (iv) & $\mathcal{M}_{++}^{x/y}$ &  & $\mathcal{R}_{-+}^{z}$ & \tabularnewline
CII & -1 & -1 & 1 & (i) & $\mathcal{M}_{++}^{x/y}$ &  & $\mathcal{R}_{+-}^{z}$* & \tabularnewline
 &  &  &  & (iv) & $\mathcal{M}_{\pm\pm}^{x/y}$ & $\mathcal{R}_{--}^{x/y}$ & $\mathcal{R}_{\pm\mp}^{z}$ & $\mathcal{I}_{-+}$\tabularnewline
C & 0 & -1 & 0 & (i) & $\mathcal{T}^{-}\mathcal{M}_{+}^{x/y}$ &  &  & \tabularnewline
 &  &  &  & (iv) & $\mathcal{T}^{\mp}\mathcal{M}_{\pm}^{x/y}$ & $\mathcal{R}_{-}^{x/y}$ &  & \tabularnewline
CI & 1 & -1 & 1 & (i) & $\mathcal{M}_{--}^{x/y}$ &  &  & \tabularnewline
 &  &  &  & (iv) & $\mathcal{M}_{\pm\pm}^{x/y}$ & $\mathcal{R}_{+-}^{x/y}$ & $\mathcal{R}_{\mp\pm}^{z}$ & $\mathcal{I}_{++}$\tabularnewline
\end{tabular}
\caption{ \label{tab:classification}
Combinations of tenfold-way symmetries ${\cal T}$, ${\cal P}$, or ${\cal C} = {\cal P}{\cal T}$ and order-two crystalline symmetries $\mathcal{S}=\mathcal{M}^{x/y}$, ${\cal R}^{x/y}$, $\mathcal{R}^z$, $\mathcal{I}$ or magnetic symmetries ${\cal T} {\cal S}$ allowing for a second-order topological gapless phase with anomalous hinge states around $k_z = 0$ (of type (i), (ii), or (iii), as discussed in the text) or with flat hinge arcs at generic $k_z$ (type (iv)). The subscripts  ${\cal S}_{\pm}$ indicate, whether ${\cal S}$ commutes ($+$) or anticommutes ($-$) with ${\cal T}$, ${\cal P}$, and/or ${\cal C}$, using the convention that the representation matrix $U_{\cal S}^2 = 1$. Magnetic symmetries are indicated by the product ${\cal T} {\cal S}$, where the superscript ${\cal T}^{\pm}$ indicates the square $({\cal T S})^2 = \pm 1$ and the subscript ${\cal S}_{\pm}$ indicates, whether ${\cal T S}$ commutes or anticommutes with ${\cal P}$ or ${\cal C}$. In class CII with ${\cal R}^z_{+-}$, the zero-energy hinge arc ends at a surface Dirac cone, see App.\ \ref{app:c} for details.}
\end{table}

\section{Classification}
In Table \ref{tab:classification}, we list all combinations of tenfold-way and crystalline symmetries for which second-order nodal semimetal or superconductor phases exist, for the twofold rotations ${\cal R}^x$ or ${\cal R}^y$, ${\cal R}^z$, mirror ${\cal M}^x$ or ${\cal M}^y$, or inversion symmetry ${\cal I}$. We include magnetic crystalline symmetries, which are the product of a crystalline symmetry ${\cal S}$ and time-reversal ${\cal T}$.
The subscripts $\pm$ in the table indicate, whether the crystalline symmetry ${\cal S}$ commutes or anticommutes with the tenfold-way symmetries ${\cal T}$, ${\cal P}$, or ${\cal C}$, if applicable (also see App.\ \ref{app:d}).
Of the symmetry classes shown in the table, a second-order semimetal with a flat arc of zero-energy hinge states was previously proposed in Ref.\ \onlinecite{wang2020a} in the presence of the magnetic inversion symmetry ${\cal T}{\cal I}_-$.
A list of second-order gapless phases protected by crystalline antisymmetries is given in App.\ \ref{app:a}.

The results presented in Table \ref{tab:classification} have been obtained by comparison of the order-resolved classifications at $k_z = 0$ and $0 < |k_z| < \pi$, as explained in detail in App.\ \ref{app:a}. 
 We supply an exhaustive list of topological invariants for the nodal phases in App.\ \ref{app:f}.
 More details on the construction and notation of the symmetry classes can be found in App.\ \ref{app:d}.
 Furthermore, in App.\ \ref{app:e} we discuss two boundary signatures that do not occur in the presence of a single twofold crystalline symmetry.
 
\section{Nodal loops vs.\ Dirac points}
For the order-two crystalline symmetries considered here, a bulk band structure with a ``Dirac point'', a multiply degenerate Weyl point, is unstable to a splitting into
a nodal loop.
This is the case, {\it e.g.}, in the model considered in Ref.\ \cite{wang2020a}, as well as for the models considered here. Nodal loops for the four examples discussed here are shown in App.\ \ref{app:b}. Stabilizing Dirac points requires crystalline symmetries of higher order than considered here, such as fourfold or sixfold rotation symmetry. Topological semimetals with flat or approximately flat arcs of hinge states and fourfold rotation symmetry (our type (i)) were discussed in Refs.\ \onlinecite{ezawa2018,lin2018semimetal,ghorashi2020,wieder2020}. Reference \onlinecite{zhang2020} proposes models for a nodal superconductor with inversion-symmetry-protected helical hinge states around $k_z = 0$ (our type (iii)) and nodal points in the bulk band structure protected by fourfold and sixfold rotation symmetry. In principle, degenerate Weyl points may also split into single Weyl points if the classification group at finite $k_z$ admits both first-order and second-order or obstructed atomic-limit phases under taking the direct sum. Such a scenario does not occur, however, for the order-two crystalline symmetries \cite{trifunovic2019}.

\section{Hinge disorder}
As long as the bulk density of states vanishes at the nodal point, zero-energy hinge arcs are robust to weak disorder \cite{szabo2020,zhang2021}. However, potential disorder may broaden a flat hinge arc, even if it respects the tenfold-way symmetries ${\cal T}$, ${\cal P}$, and ${\cal C}$ (if applicable). Such local hybridization of hinge states is absent, however, if the hinge states are protected by a chiral antisymmetry ${\cal C}$ and all hinge states have the same ${\cal C}$-parity. This is the case for our example \eqref{eq:H_AIII_My+}. If the hinge states appear in pairs with opposite ${\cal C}$-parity or if they are not pinned to zero energy by an antisymmetry ${\cal C}$, no such protection exists. In our examples \eqref{eq:H_DIII_M--} and \eqref{eq:H_DIII_Rz-+}, the zero-energy hinge states appear in Kramers pairs with opposite ${\cal C}$-parity, allowing them to hybridize. For type (iii), the hinge states are chiral or helical Majorana modes, which are prohibited from localization by disorder, although away from zero energy disorder may cause chiral or helical hinge states to hybridize with bulk states. A detailed discussion on the effect of hinge disorder for each example discussed in this manuscript can be found in App.\ \ref{app:dis}.

\section{Conclusion}
Using the systematic relation between topological gapless phases in $d$ dimensions and topological gapped phases in $d-1$ dimensions, we performed an extensive topological classification for $d=3$ and identified four types of second-order topological semimetals and nodal superconductors distinguished by their boundary signatures and associated bulk topology around the nodal manifold. 
 These boundary signatures are unique to nodal topological phases as the hinge anomaly requires a bulk node before connecting to a topologically trivial configuration at large momentum.
For a strict protection, all types require the presence of an antisymmetry, which suggests superconductors or semimetals with an approximate sublattice antisymmetry as natural candidates. 
We hope that our work may contribute to the topological interpretation of nodal superconductors, such as doped Bi$_2$Se$_3$ \cite{smylie2016, smylie2017, smylie2018} where the proposed multicomponent order parameter with $E_u$ symmetry can yield a system in class DIII with mirror $\mathcal{M}_{--}$ and rotation $\mathcal{R}_{-+}$ permitting second-order nodal topology ({\it c.f.} Table I), as well as iron-based \cite{fletcher2009, paglione2010, yusuke2010, hashimoto2010, zhang2012, hashimoto2012, su2012, hong2013, kim2013, johnson2015, smidman2018} and heavy-fermion compounds \cite{ikeda2015} where nodal loops have been observed.

We thank Luka Trifunovic and Michele Burrello for discussions.
This research was supported by the Deutsche Forschungsgemeinschaft (DFG, German Research Foundation) - Project Number 277101999 - CRC 183 (projects A02, A03, and C01). MG acknowledges support by the European Research Council (ERC) under the European Union's Horizon 2020 research and innovation program under grant agreement No.~856526, and from the Danish National Research Foundation, the Danish Council for Independent Research $\vert$ Natural Sciences.

\appendix

\section{Further discussion of examples from the main text} 
\label{app:b}

\begin{figure*}
\includegraphics[width=2\columnwidth]{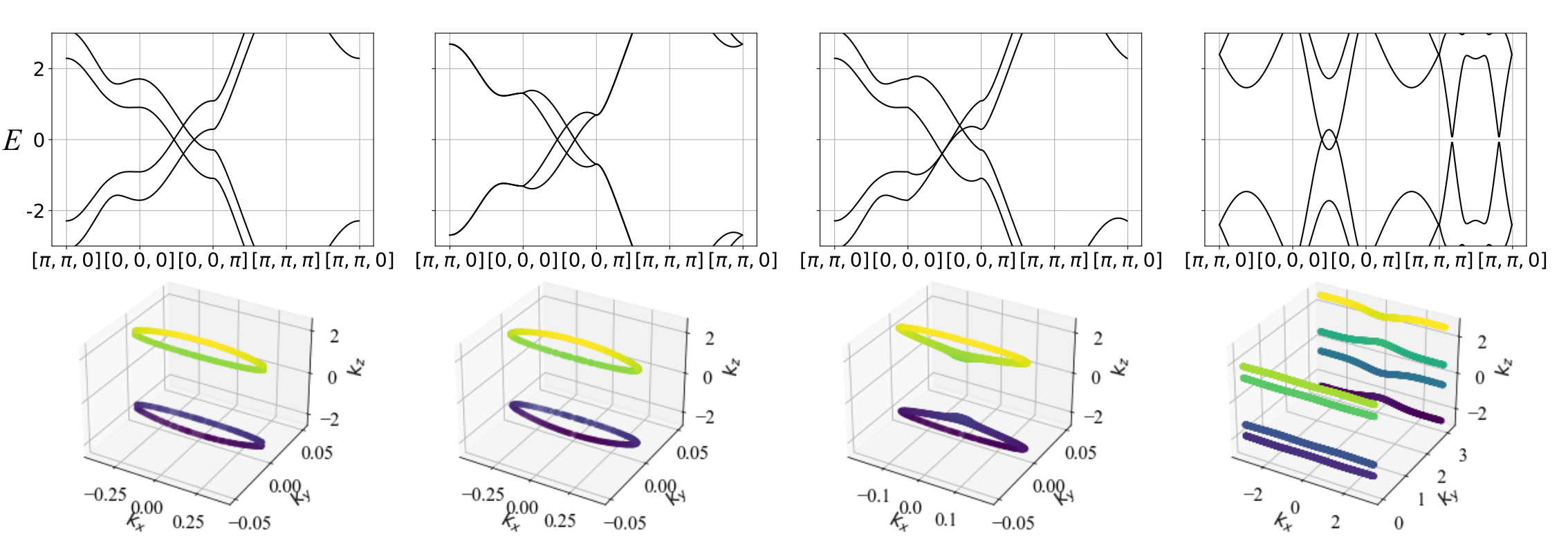}
\caption{\label{fig:bandstructures_bulk}
Bulk band structures (top) and nodal lines in the three-dimensional Brillouin zone (bottom) for the representative models of Eqs. \eqref{eq:H_AIII_My+}, \eqref{eq:H_DIII_M--}, \eqref{eq:H_D_Rz-}, and \eqref{eq:H_DIII_Rz-+} (from left to right) from the main text. The same parameters were used as in Fig.~\ref{fig:bandstructures}. In the depictions of the nodal lines, we include a color gradient proportional to the $k_z$ coordinate of the nodal line.}
\end{figure*}

To decide, whether a certain combination of tenfold-way and crystalline symmetries can protect a second-order topological semimetal phase and to find the type of the boundary signatures, one needs to
\begin{itemize}
\item[(a)] determine the crystalline symmetry classes of the two-dimensional Hamiltonian $H_{k_z}(k_x,k_y)$ at the high-symmetry planes $k_z = 0$, $\pi$, and at generic $0 < |k_z| < \pi$, and
\item[(b)] obtain the order-resolved classification corresponding to the crystalline symmetry classes at $k_z=0$, $\pi$, and at generic $0 < |k_z| < \pi$, together with the mapping between these.
\end{itemize}
For the twofold crystalline symmetries, order-resolved classification information can be found in Ref.\ \onlinecite{trifunovic2019}, which classifies two-dimensional topological band structures in terms of a ``subgroup sequence''
$$
K'' \subseteq K' \subseteq K.
$$%
Here $K$ is the full classifying group (restricted to strong topological phases and subject to the rules of stable equivalence \cite{chiu2016}), $K'$ classifies all topological band structures that do not have first-order boundary signatures, and $K''$ classifies all topological band structures without boundary signatures. The quotient $K'/K''$ then describes the second-order phases. Since the crystalline symmetry class at generic $0 < |k_z| < \pi$ is obtained from the crystalline symmetry class at the high-symmetry planes $k_z = 0$, $\pi$ by lifting symmetry requirements, in the mathematical literature the mapping between the classifying groups at $k_z = 0$, $\pi$ and at generic $0 < |k_z| < \pi$ is known as the ``forgetful functor'' \cite{chiu2016}. To the best of our knowledge, no complete classification information that includes the action of the forgetful functor for the tenfold-way classes with a twofold crystalline symmetry is available in the literature, but this information can be obtained relatively straightforwardly by inspecting models for the generators of the classifying groups.

We now discuss the crystalline symmetry classes of the two-dimensional Hamiltonian $H_{k_z}(k_x,k_y)$ at the high-symmetry planes $k_z = 0$, $\pi$, and at generic $0 < |k_z| < \pi$, the order-resolved classifications, the action of the forgetful functor, and the resulting consequences for three-dimensional topological gapless phases for the four examples discussed in the main text. 
Fig.~\ref{fig:bandstructures_bulk} shows the band structure and the nodal loops calculated from the model Hamiltonians with perturbation.

{\it Example of Eq.\ (\ref{eq:H_AIII_My+}). --- } Considering $k_z$ as a parameter, the three-dimensional Hamiltonian $H(k_x,k_y,k_z)$ of Eq.\ (\ref{eq:H_AIII_My+}) describes a one-parameter family of two-dimensional Hamiltonians $H_{k_z}(k_x,k_y)$. At the high-symmetry planes $k_z = 0$ and $k_z = \pi$, $H_{k_z}(k_x,k_y)$ is subject to the chiral antisymmetry ${\cal C}$ and the mirror symmetry ${\cal M}^x$ as
\begin{align}
  H_{0,\pi}(k_x,k_y) =&\, - U_{\cal C} H_{0,\pi}(k_x,k_y) U_{\cal C}^{\dagger} \nonumber \\ = &\,
  U_{\cal M} H_{0,\pi}(-k_x,k_y) U_{\cal M}^{\dagger},
\end{align}
with $U_{\cal C} = \tau_3 \sigma_0$ and $U_{\cal M} = \tau_3 \sigma_3$, see Eq.\ (\ref{eq:symm_AIII_My+}) of the main text. In the notation of Ref.\ \onlinecite{trifunovic2019}, this is the crystalline symmetry class $\mbox{AIII}^{{\cal M}_+}$, where the Cartan symbol ``AIII'' indicates the presence of the chiral antisymmetry ${\cal C}$ and the superscript ``${\cal M}_+$'' the presence of the crystalline mirror symmetry ${\cal M}^x$ that commutes with ${\cal C}$. Since the mirror symmetry leaves $z$ invariant, the crystalline symmetry class for generic $0 < |k_z| < \pi$ is the same.

The subgroup sequence classifying the crystalline symmetry class $\mbox{AIII}^{{\cal M}_+}$ is given in Table II of Ref.\ \cite{trifunovic2019},
$$
0 \subseteq \ZZ \subseteq \ZZ,
$$%
from which one immediately concludes that all nontrivial topological phases in the crystalline symmetry class $\mbox{AIII}^{{\cal M}_+}$ are second-order phases with zero-energy corner states at mirror-symmetric corners. The four-band Hamiltonian (\ref{eq:H_AIII_My+}) is a generator of the nontrivial topology. One can see this, {\it e.g.}, by noting that it has the maximal number of anticommuting terms and the minimal number of bands. Alternatively, that the four-band Hamiltonian (\ref{eq:H_AIII_My+}) with $|k_z| < k_0$ is a generator of the nontrivial topology can also be inferred from the exact diagonalization, which gives precisely one zero-energy corner state per mirror-symmetric corner if $|k_z| < k_0$, see Fig.\ \ref{fig:bandstructures}(b) in the main text. 

{\it Example of Eq.\ (\ref{eq:H_DIII_M--}). --- } When seen as a $k_z$-dependent family of two-dimensional Hamiltonians $H_{k_z}(k_x,k_y)$, the model (\ref{eq:H_DIII_M--}) is in symmetry class $\mbox{DIII}^{{\cal M}_{--}}$ at the high-symmetry planes $k_z = 0$, $\pi$. For generic $0 < |k_z| < \pi$, $H_{k_z}(k_x,k_y)$ is constrained by the product ${\cal C} = {\cal P} {\cal T}$ and by the mirror symmetry ${\cal M}$ only. Since ${\cal M}$ commutes with ${\cal C}$, $H_{k_z}(k_x,k_y)$ is in the symmetry class $\mbox{AIII}^{{\cal M}_+}$ for generic $0 < |k_z| < \pi$. The subgroup sequences for $k_z = 0$, $\pi$ and for generic $0 < |k_z| < \pi$ are given in Tables IV and II of Ref.\ \onlinecite{trifunovic2019}, respectively,
\begin{align}
  \begin{array}{ll}
  0 \subseteq 2 \ZZ \subseteq \ZZ & \mbox{for $k_z = 0$, $\pi$}, \\
  0 \subseteq \ZZ \subseteq \ZZ & \mbox{for $0 < |k_z| < \pi$}.
  \end{array}
  \label{eq:subgroupSi}
\end{align}
{}From these subgroup sequences we conclude that the generator of the classifying group is a first-order phase for $k_z = 0$, $\pi$ and a second-order phase for $0 < |k_z| < \pi$. That the four-band Hamiltonian (\ref{eq:H_DIII_M--}) is a generator for $k_z = 0$ as well as $0 < |k_z| < k_0$ can be seen by noting that it has the maximal number of anticommuting terms and the minimal number of bands or, alternatively, by inspection of the exact diagonalization results of Fig.\ \ref{fig:bandstructures}(b), which show that the two-dimensional Hamiltonian $H_{k_z}(k_x,k_y)$ corresponding to Eq.\ (\ref{eq:H_DIII_M--}) has helical Majorana edge states for $k_z = 0$ and non-degenerate zero-energy corner states for $0 < k_z < k_0$.

{\it Example of Eq.\ (\ref{eq:H_D_Rz-}). --- } The model of Eq.\ (\ref{eq:H_D_Rz-}) satisfies the symmetry constraints
\begin{align}
  H(k_x,k_y,k_z) =&\, -U_{\cal P} H(-k_x,-k_y,-k_z)^* U_{\cal P}^{\dagger} \nonumber \\ =&\, U_{\cal R} H(-k_x,-k_y,k_z) U_{\cal R}^{\dagger},
\end{align}
with $U_{\cal P} = \tau_1 \sigma_0$ and $i U_{\cal R} = i \tau_3 \sigma_3$. It is in the crystalline symmetry classes $\mbox{D}^{{\cal R}_-}$ for $k_z = 0$, $\pi$ and $\mbox{A}^{{\cal R}}$ for $0 < |k_z| < \pi$. The corresponding subgroup sequences are \cite{trifunovic2019}
$$
  \begin{array}{ll}
  2 \ZZ \subseteq \ZZ \subseteq \ZZ^2 & \mbox{for $k_z = 0$, $\pi$}, \\
  \ZZ \subseteq \ZZ \subseteq \ZZ^2 & \mbox{for $0 < |k_z| < \pi$}.
  \end{array}
$$%
The two-dimensional Hamiltonian $H_{k_z}(k_x,k_y)$ corresponding to the model (\ref{eq:H_D_Rz-}) is the generator ``1'' of $K'$ for $|k_z| < k_0$, which is a second-order phase with zero-energy corner states for $k_z = 0$ and an obstructed atomic-limit phase for $0 < |k_z| < k_0$. At $|k_z| = k_0$ there is a transition between two atomic-limit phases, which lends topological protection to the gaplessness, but does not come with a boundary signature. This explains the phenomenology of the three-dimensional model (\ref{eq:H_D_Rz-}): dispersing chiral hinge modes around $k_z = 0$, pinned to zero energy at $k_z = 0$.
  
{\it Example of Eq.\ (\ref{eq:H_DIII_Rz-+}). --- }
The model of Eq.\ (\ref{eq:H_DIII_Rz-+}) satisfies the symmetry constraints
\begin{align}
  H(k_x,k_y,k_z) =&\, -U_{\cal P} H(-k_x,-k_y,-k_z)^* U_{\cal P}^{\dagger} \nonumber \\ =&\, U_{\cal T} H(-k_x,-k_y,-k_z)^* U_{\cal T}^{\dagger} \nonumber \\ =&\, U_{\cal R} H(-k_x,-k_y,k_z) U_{\cal R}^{\dagger},
\end{align}
with $U_\mathcal{T} = \tau_0 \sigma_2$, $U_\mathcal{P} = \tau_1 \sigma_0$, and $i U_{\mathcal{R}} = i \tau_0 \sigma_3$. At the high-symmetry momenta $k_z= 0$, $\pi$, the two-dimensional Hamitonian $H_{k_z}(k_x,k_y)$ corresponding to this model is in the crystalline symmetry class $\mbox{DIII}^{{\cal R}_{-+}}$; for generic momenta $0 < |k_z| < \pi$, the crystalline symmetry class is $\mbox{AIII}^{{\cal R}_-}$. The corresponding subgroup sequences are \cite{trifunovic2019}
$$
  \begin{array}{ll}
  0 \subseteq \ZZ_2 \subseteq \ZZ_2 & \mbox{for $k_z = 0$, $\pi$}, \\
  2 \ZZ \subseteq \ZZ \subseteq \ZZ & \mbox{for $0 < |k_z| < \pi$}.
  \end{array}
$$%
Although the symmetries allow for a nontrivial topological phase at $k_z = 0$, the corresponding Hamiltonian immediately turns trivial upon going to nonzero $k_z$, ruling out a topological semimetal phase that derives its protection from a non-trivial phase of $H_{k_z}(k_x, k_y)$ at $k_z = 0$. (This observation, as well as the converse, that any topologically non-trivial gapped two-dimensional Hamiltonian $H_{k_z}(k_x,k_y)$ must be trivial for $k_z \to 0$, $\pi$ follows from the observation that there is no nontrivial map from $\ZZ_2$ to $\ZZ$.) Instead, the family of two-dimensional Hamiltonians $H_{k_z}(k_x,k_y)$ as defined from Eq.\ (\ref{eq:H_DIII_Rz-+}) corresponds to the trivial class for $k_z = 0$ and $\pi$ --- {\it i.e.}, it corresponds to the element ``0'' in the classifying group $K$ --- and is nontrivial --- corresponding to the element ``1'' --- in a region $k_{01} < |k_z| < k_{02}$ around $|k_z| = \pi/2$. The resulting phenomenology of the three-dimensional model is as shown in Fig.\ \ref{fig:bandstructures}: flat arcs of zero-energy hinge states in a finite region around $|k_z| = \pi/2$, but no boundary signatures at $k_z = 0$ or $\pi$. 

\section{More examples} 
\label{app:c}

Below we discuss more examples that realize variants of the four main scenarios for second-order gapless topological band structures discussed in the main text.

\begin{figure*}
\includegraphics[width=2\columnwidth]{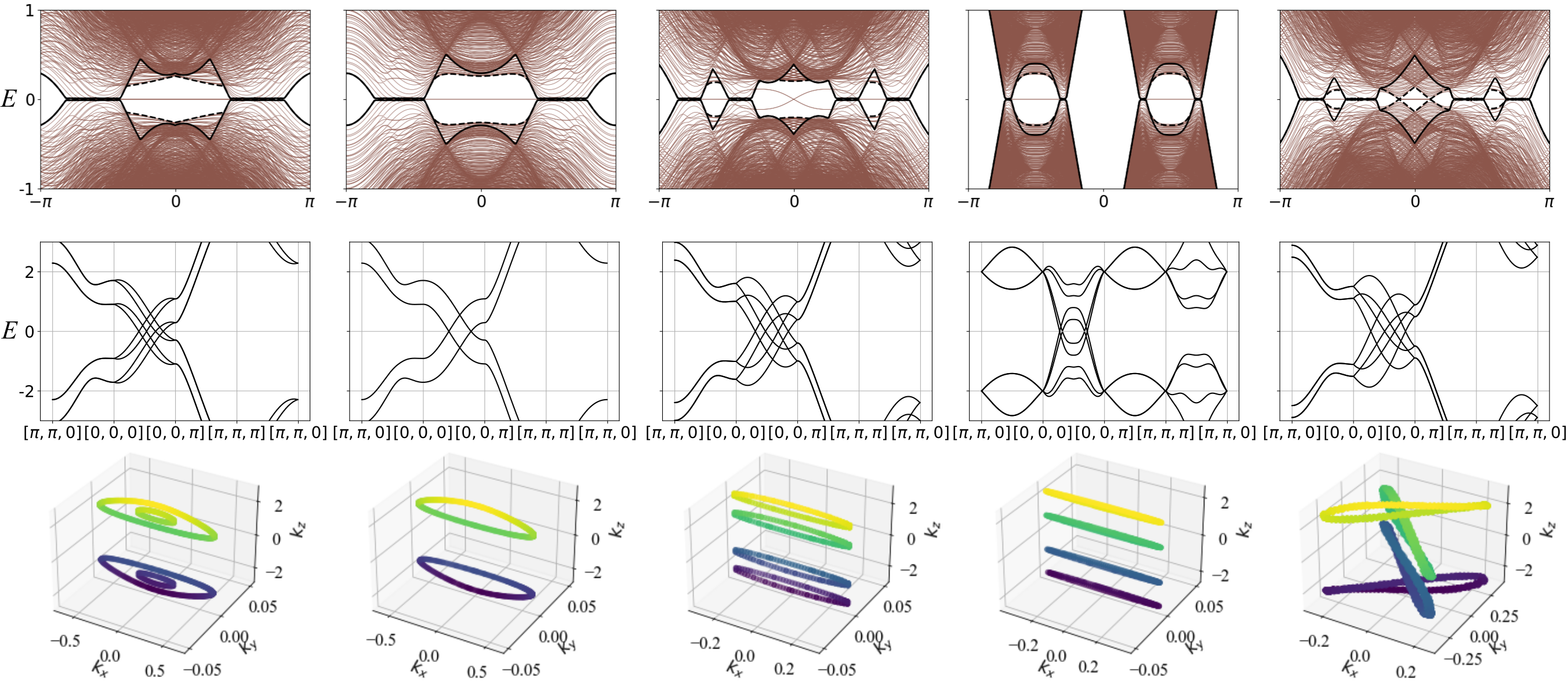}
\caption{\label{fig:app_bandstructures}
Band structures in a rhombic pillar geometry for the example systems as defined in Eqs. \eqref{eq:H_DIII_M--_type1}, \eqref{eq:H_D_Ry+}, \eqref{eq:H_DIII_Rz+-}, \eqref{eq:H_DIII_Rz-+_8bands}, and \eqref{eq:H_CII_Rz+-} (from left to right, top) and bulk band structures (middle) and nodal line (bottom). In the pillar geometry (top), periodic boundary conditions are applied in the $z$ direction, which is along the pillar axis. The full and dashed black curves denote the excitation gap in the bulk and on the surface, respectively. Symmetry-preserving perturbations have been added to the models given in the text to turn accidentally degenerate nodal points into nodal loops and to gap out non-anomalous surface states, as desribed the text. In the models of Eqs.\ \eqref{eq:H_DIII_M--_type1}, \eqref{eq:H_D_Ry+},  \eqref{eq:H_DIII_Rz+-}, and \eqref{eq:H_CII_Rz+-} we took the parameter values $t=1$, $k_0 = 0.6\pi$; for the model of Eq.\ \eqref{eq:H_DIII_Rz-+_8bands} we took the parameter values $m=0$, $t=1$, $t'=2$}
\end{figure*}

{\it Type (i) with degenerate flat hinge arcs. ---} The example of Eq.\ (\ref{eq:H_AIII_My+}) describes a second-order topological gapless phase with non-degenerate flat arcs of hinge states. Since the hinge states are protected by the chiral antisymmetry ${\cal C}$ and, hence, have a $\ZZ$ topological classification, examples with degenerate flat arcs of hinge states can trivially be obtained by taking the direct sum of multiple copies of the model (\ref{eq:H_AIII_My+}). A nontrivial example with twofold degenerate hinge arcs, for which the degeneracy at $k_z = 0$ is symmetry-enforced, can be found by considering the direct sum of two copies of the example of Eq.\ (\ref{eq:H_DIII_M--}) of the main text, 
\begin{align}
 H(k_x,k_y,k_z) = & \, t \tau_3 \sigma_0 \rho_0 [m(k_z) + 2 - \cos k_x - \cos k_y] \nonumber \\
 &\, \mbox{} + t \tau_2 \sigma_0 \rho_0 \sin k_x + t \tau_1 \sigma_3 \rho_0 \sin k_y.
\label{eq:H_DIII_M--_type1}
\end{align}
Here, the $2 \times 2$ identity matrix $\rho_0$ and the associated Pauli matrices $\rho_j$, $j=1,2,3$, describe an additional two-component degree of freedom. The tenfold-way and crystalline symmetries of this eight-band model are the same in the example of Eq.\ (\ref{eq:H_DIII_M--}) of the main text, see the previous Section for details. As before, we consider the model (\ref{eq:H_DIII_M--_type1}) as a one-parameter family of two-dimensional Hamiltonians $H_{k_z}(k_x,k_y)$. Because the model (\ref{eq:H_DIII_M--_type1}) is the direct sum of two copies of a Hamiltonian that generates the classification group $K$, it corresponds to the element ``2'' in the sequence of classifying groups.  According to the subgroup sequences (\ref{eq:subgroupSi}), the element ``2'' of the classying group is a second-order phase with Kramers-degenerate zero-energy corner states for $k_z = 0$ and a second-order phase with twofold degenerate corner states for $0 < |k_z| < k_0$. The resulting three-dimensional band structure has a flat arc of twofold degenerate zero-energy hinge modes for $-k_0 < k_z < k_0$, whereby the twofold degeneracy at $k_z = 0$ is protected by time-reversal symmetry, whereas the twofold degeneracy at $0 < |k_z| < k_0$ is protected by the chiral antisymmetry ${\cal C} = {\cal P} {\cal T}$.

Figure \ref{fig:app_bandstructures} shows the nodal region, the bulk band structure, and the dispersion for a pillar geometry with translation invariance in the $z$ direction. To remove non-topological surface states and accidental degeneracies in the spectrum, the additional term
\begin{equation}
  \delta H = b_1 \tau_0 \sigma_1 \rho_2 + b_2 \tau_0 \sigma_1 \rho_0 \sin k_z 
  \label{eq:Hadd2}
\end{equation}
has been added to the Hamiltonian (\ref{eq:H_DIII_M--_type1}), with parameter values $b_1 = 0.4t$ and $b_2 = 0.2t$. 

Although this model has the smallest number of bands compatible with the requirement that it has Kramers degenerate hinge modes at $k_z = 0$, it is not a minimal model. In the context of the classification, this follows from the fact that the model (\ref{eq:H_DIII_M--_type1}) corresponds to the non-minimal element ``2'' in the classifying group. It also follows from the phenomenology of the model: In principle, one can add a perturbation that preserves time-reversal symmetry, but causes the transition of $H_{k_z}(k_x,k_y)$ from the nontrivial phase ``2'' at $k_z = 0$ towards the trivial phase ``0'' at $k_z = \pi$ to proceed in two steps. As a result, there are nodal points at two values of $k_z$, each of which can be broadened into a nodal loop by a symmetry-preserving perturbation. For $k_z$ between the projections of the nodal points/loops there are non-degenerate zero-energy hinge states, as in the original model (\ref{eq:H_DIII_M--}) of the main text.

{\it Type (i) with rotation symmetry $\mathcal{R}^y$. ---}
Another example for a second-order nodal superconductor with a flat arc of zero-energy hinge states around $k_z = 0$ is a time-reversal symmetry breaking superconductor with a twofold rotation symmetry ${\cal R}^y$ around the $y$ axis and even-parity superconductivity. A concrete lattice model is the four-band model
\begin{align}
  H(k_x,k_y,k_z) = & \, t  \tau_3 \sigma_0 (m(k_z) + 2 - \cos k_x - \cos k_y)
  \nonumber \\
 & + t \tau_1 \sigma_1 \sin k_x + t \tau_1 \sigma_3 \sin k_y,
\label{eq:H_D_Ry+}
\end{align}
which satisfies the symmetry constraints
\begin{align}
  H(k_x,k_y,k_z) =&\, -U_{\cal P}  H(-k_x,-k_y,-k_z)^* U_{\cal P}^{\dagger} \nonumber \\ =&\,
  U_{\cal R} H(-k_x,k_y,-k_z) U_{\cal R}^{\dagger},
\end{align}
with $U_\mathcal{P} = \tau_1 \sigma_0$ and $i U_{\mathcal{R}} = i \tau_0 \sigma_3$. Since the twofold rotation ${\cal R}^y$ acts like a mirror $k_x \to -k_x$ at the high-symmetry values $k_z = 0$ or $k_z = \pi$, the crystalline symmetry class of the two-dimensional Hamiltonian $H_{k_z}(k_x,k_y)$ at $k_z = 0$, $\pi$ is $\mbox{D}^{{\cal M}_+}$. At generic $k_z$, $H_{k_z}(k_x,k_y)$ is constrained by the product ${\cal P}{\cal R}$ only, which, at a fixed value of $k_z$, effectively acts as an antiunitary mirror antisymmetry ${\cal P}{\cal M}^x$, so that $H_{k_z}(k_x,k_y)$ is in the crystalline symmetry class $\mbox{A}^{{\cal P}^+{\cal M}}$. From Tables III and IV of Ref.\ \onlinecite{trifunovic2019} we then find that in both cases the order-resolved topological classifications given by the subgroup sequence
$$
  0 \subseteq \ZZ_2 \subseteq \ZZ_2.
$$%
According to this subgroup sequence, the only topologically nontrivial band structures are of second order. For $|k_z| < k_0$ the model (\ref{eq:H_D_Ry+}) realizes such a topological band structure, for $k_z = 0$ as well as $0 < |k_z| < k_0$. It has zero-energy hinge states for $-k_0 < k_z < k_0$ on hinges that are mapped to themselves under the twofold rotation symmetry ${\cal R}^y$.
  
To remove non-topological surface states and accidental degeneracies, for the numerical calculations we add the additional term
\begin{equation}
  \delta H =  b_1 \tau_3 \sigma_3 + b_2 \tau_0 \sigma_1 \sin k_z
\end{equation}
with $b_1 = b_2 = 0.4t$. The corresponding bulk band structure and the quasi-one-dimensional band structure of a rhombic pillar geometry are shown in Fig.~\ref{fig:app_bandstructures}. 

{\it Type (ii) with inversion symmetry. --- }
The model of Eq.\ (\ref{eq:H_DIII_M--}) of the main text, which describes an odd-parity nodal superconductor with mixed-order boundary signatures (surface Dirac cones pinned to $k_z = 0$ and flat arcs of hinge states around $k_z = 0$) also satisfies an unusual inversion symmetry $\mathcal{I}$ with representation $U_\mathcal{I} = \tau_3 \sigma_3$. (This inversion symmetry is unusual, because it anticommutes with time reversal.) If ${\cal I}$ is imposed instead of ${\cal M}^x$, the model of Eq.\ (\ref{eq:H_DIII_M--}) is still topological, with crystalline symmetry class $\mbox{DIII}^{{\cal R}_{--}}$ for $k_z = 0$, $\pi$ and $\mbox{AIII}^{{\cal T}^+{\cal R}_-}$ for $0 < |k_z| < \pi$. (Note that at fixed $k_z = 0$, $\pi$, inversion effectively acts as a twofold rotation ${\cal R}^z$. The same applies to the product ${\cal T} {\cal I}$ at fixed $0 < |k_z| < \pi$.) The subgroup sequences follow from Tables IV and III of Ref.\ \onlinecite{trifunovic2019},
$$
  \begin{array}{ll}
  0 \subseteq \ZZ_2 \subseteq \ZZ_2^2 & \mbox{for $k_z = 0$, $\pi$}, \\
  0 \subseteq \ZZ_2 \subseteq \ZZ_2 & \mbox{for $0 < |k_z| < \pi$}.
  \end{array}
$$%
Hence, at $k_z = 0$, $\pi$, there are first-order as well as second-order phases, with separate $\ZZ_2$ classifications. At generic $0 < |k_z| < \pi$ there is only a second-order phase. Upon continuously going from $k_z = 0$ to generic $0 < |k_z| < \pi$, the generator of the first-order phase at $k_z = 0$ becomes a second-order phase at nonzero $k_z$. That this is the case is seen by inspecting the behavior of $H_{k_z}(k_x,k_y)$ for the concrete model (\ref{eq:H_DIII_M--}), which has first-order boundary states at $k_z = 0$ and zero-energy corner states for $0 < k_z < k_0$ (see discussion in the main text and in the previous Section). Hence, we conclude that the model (\ref{eq:H_DIII_M--}) with inversion symmetry ${\cal I}_{--}$ instead of the mirror symmetry ${\cal M}^x$ also realizes a second-order nodal superconductor of type (ii). A difference between the inversion-symmetric and mirror-symmetric versions of the model (\ref{eq:H_DIII_M--}) is that in the mirror-symmetric version the hinge states appear on the mirror-symmetric hinges, whereas in the inversion-symmetric version the hinge states can appear on any pair of inversion-related hinges. Since the phenomenology of the inversion-symmetric version of the model (\ref{eq:H_DIII_M--}) is the same as that of the mirror-symmetric version, we refer to Fig.\ \ref{fig:bandstructures}(b) of the main text for numerical results.

{\it Type (iii) with helical hinge modes. ---}
A second-order topological nodal superconductor of type (iii) with helical hinge modes exists in the (three-dimensional) symmetry class $\mbox{DIII}^{\mathcal{R}_{+-}^z}$, which describes an odd-rotation-parity time-reversal-symmetric superconductor with an unconventional twofold rotation symmetry ${\cal R}^z$. (The rotation symmetry is unconventional, because ${\cal R}^2 = 1$ and ${\cal R}$ commutes with time reversal.) To construct a concrete lattice model, one may take the direct sum $H(k_x,k_y,k_z) \oplus H^*(-k_x,k_y,k_z)$ of two time-reversed copies of the four-band model $H(k_x,k_y,k_z)$ of Eq. \eqref{eq:H_D_Rz-},
\begin{align}
   H(k_x,k_y,k_z) = & \, t \tau_3 \sigma_3 \rho_0 [m(k_z) + 2 - \cos k_x - \cos k_y] \nonumber \\
   &\, \mbox{} + t \tau_1 \sigma_3 \rho_3 \sin k_x + t \tau_2 \sigma_3 \rho_0 \sin k_y,
\label{eq:H_DIII_Rz+-}
\end{align}
where the Pauli matrices $\rho_j$ describe an additional two-component degree of freedom. The model has the symmetries
\begin{align}
  H(k_x,k_y,k_z) =&\, -U_{\cal P}  H(-k_x,-k_y,-k_z)^* U_{\cal P}^{\dagger} \nonumber \\ =&\, U_{\cal T}  H(-k_x,-k_y,-k_z)^* U_{\cal T}^{\dagger} \nonumber \\ =&\,
  U_{\cal R} H(-k_x,-k_y,k_z) U_{\cal R}^{\dagger},
\end{align}  
with $U_\mathcal{P} = \tau_1 \sigma_0 \rho_0$, $U_\mathcal{T} = \tau_0 \sigma_0 \rho_2$, and $U_{\mathcal{R}^z} = \tau_3 \sigma_3 \rho_0$. For the high-symmetry momenta $k_z = 0$, $\pi$, the two-dimensional Hamiltonian $H_{k_z}(k_x,k_y)$ corresponding to Eq.\ (\ref{eq:H_DIII_Rz+-}) is in crystalline symmetry class $\mbox{DIII}^{{\cal R}_{+-}}$; and at generic $0 < |k_z| < \pi$ it is in symmetry class $\mbox{AIII}^{{\cal R}_{-}}$. The corresponding subgroup sequences follow from Tables IV and II of Ref.\ \onlinecite{trifunovic2019},
$$
  \begin{array}{ll}
  4 \ZZ \subseteq 2 \ZZ \subseteq \ZZ & \mbox{for $k_z = 0$, $\pi$}, \\
  2 \ZZ \subseteq \ZZ \subseteq \ZZ & \mbox{for $0 < |k_z| < \pi$}.
  \end{array}
$$%
For $k_z = 0$ the eight-band model (\ref{eq:H_DIII_Rz+-}) corresponds to the element ``2'' in this subgroup sequence, which corresponds to a second-order band structure with Kramers pairs of zero-energy corner states. For $0 < k_z < k_0$ the element ``2'' in the subgroup sequence corresponds to a topologically nontrivial atomic limit.

Like the model of Eq.\ (\ref{eq:H_DIII_M--_type1}), the model (\ref{eq:H_DIII_Rz+-}) is not a minimal model. As a function of $k_z$, the two-dimensional Hamiltonian $H_{k_z}(k_x,k_y)$ can transition from the nontrivial second-order phase ``2'' at $k_z=0$ to the trivial phase at $k_z = \pi$ via an intermediate gapped phase corresponding to the generator ``1'' of the classification group $K$. If the gap closings take place at $k_z = \pm k_{01}$ and $k_z = \pm k_{02}$, the three-dimensional system has helical dispersing hinge modes around $k_z = 0$ and flat arcs of zero modes for $|k_z|$ between $k_{01}$ and $k_{02}$. As for all other models considered here, a symmetry-preserving perturbation may change the gap closing points into nodal loops.

A numerical evaluation of the model (\ref{eq:H_DIII_Rz+-}), with an additional term
\begin{equation}
  \delta H = b_1 \tau_2 \sigma_2 \rho_0 + b_2 \tau_1 \sigma_1 \rho_3 \sin k_z + b_3 \tau_2 \sigma_2 \rho_2 \sin k_z
\end{equation}
to gap out non-topological surface states and to separate the gap closing points, is shown in Fig.\ \ref{fig:app_bandstructures}. The parameter values are $b_1 = b_2 = 0.3t,\ b_3 = 0.6t$. 

{\it Type (iv) with four nodal points. ---}
The model \eqref{eq:H_DIII_Rz-+} of a second-order topological superconductor in symmetry class $\mbox{DIII}^{\mathcal{R}^z_{-+}}$ has eight nodal loops. Four of the nodal loops are a consequence of a transition between obstructed atomic limits. This obstruction is ``fragile'', in the sense that these four nodal loops can be removed under the direct sum with the corresponding atomic limit that cancels the obstruction. This cancellation is realized, {\it e.g.}, in an eight-band generalization of the model of Eq. \eqref{eq:H_DIII_Rz-+},
\begin{align}
  H(k_{x},k_{y},k_{z}) =&\, t\tau_{0} \sigma_{3}
  \sin k_z \nonumber \\ &\, \ \ \mbox{} \times [\mu_3 (m-1) - \mu_{11} (\cos k_{x}  + \cos k_{y})] \nonumber \\ 
  &\, \mbox{}+ t \tau_{0} \sigma_{1} \mu_{11} \sin k_{x} + t \tau_{3} \sigma_{2} \mu_{11} \sin k_{y} \nonumber \\
  &\, \mbox{}+ t' \tau_3 \sigma_0 \mu_1 \cos^2 k_z,
\label{eq:H_DIII_Rz-+_8bands}
\end{align}
where $\mu_{11} = \frac{1}{2}(\mu_0 + \mu_3)$.
The bulk band structure, the manifold of nodal excitations in the three-dimensional Brillouin zone, and the band structure in a pillar geometry are shown in Fig. \ref{fig:app_bandstructures}. In the numerical calculations, we included an additional symmetry-allowed hybridization
\begin{equation}
  \delta H = b_1 \tau_2 \sigma_0 \mu_0 \sin k_z + b_2 \tau_2 \sigma_0 \mu_3 \sin k_z
\end{equation}
with $b_1 = 0.4t,\ b_2 = 0.2t$ to gap out non-topological surface states and remove accidental spectral degeneracies. 

{\it Type (i)* in class CII with rotation symmetry $\mathcal{R}^z_{+-}$. ---}
In symmetry class CII with rotation symmetry $\mathcal{R}^z_{+-}$, we encounter a special case of a second-order topological semimetal with zero-energy hinge arcs that, for generic $0 < |k_z| < \pi$, are only protected by the boundary gap, but not by the bulk gap. The subgroup sequences for this crystalline symmetry class are \cite{trifunovic2019}
$$
  \begin{array}{ll}
  4 \ZZ \subseteq 2 \ZZ \subseteq 2 \ZZ & \mbox{for $k_z = 0$, $\pi$}, \\
  2 \ZZ \subseteq \ZZ \subseteq \ZZ & \mbox{for $0 < |k_z| < \pi$}.
  \end{array}
$$%
  The subgroup sequence for $k_z = 0$ indicates that for this momentum, the system permits a second-order topological phase hosting an anomalous, protected Kramers pair of zero-energy corner states. In class CII, the two partners in the pair have the same eigenvalue under the chiral antisymmetry $\mathcal{C} = \mathcal{TP}$, because ${\cal T}$ commutes with ${\cal C}$ \cite{geier2018}. These hinge states must remain at zero energy upon going to nonzero $k_z$, because the chiral antisymmetry prohibits them from acquiring a non-zero energy. However, since ${\cal T}$ maps $k_z$ to $-k_z$, it no longer pairs zero-energy hinge state at nonzero $k_z$. It is possible to move and hybridize such zero-energy hinge states with the zero-energy modes at the opposite hinge, while preserving rotation symmetry \cite{geier2018}. This transformation closes only the gap on the surface, but not in the bulk. This is in agreement with the subgroup sequence for finite $k_z$, according to which the generator of the nontrivial topology at $k_z = 0$, which corresponds to the element ``2'', maps to an obstructed atomic-limit phase at nonzero $k_z$. The obstructed atomic limit $H_{k_z}(k_x,k_y)$ does not have in-gap corner states that are protected by the bulk topology.

These findings are realized in the eight-band lattice model
\begin{align}
   H(k_x,k_y,k_z) = & \, t \tau_3 \sigma_1 \rho_0 [m(k_z) + 2 - \cos k_x - \cos k_y] \nonumber \\
   &\, \mbox{} + t \tau_0 \sigma_3 \rho_3 \sin k_x + t \tau_0 \sigma_3 \rho_1 \sin k_y,
\label{eq:H_CII_Rz+-}
\end{align}
which satisfies the symmetries
\begin{align}
  H(k_x,k_y,k_z) =&\, -U_{\cal P}  H(-k_x,-k_y,-k_z)^* U_{\cal P}^{\dagger} \nonumber \\ =&\, U_{\cal T}  H(-k_x,-k_y,-k_z)^* U_{\cal T}^{\dagger} \nonumber \\ =&\,
  U_{\cal R} H(-k_x,-k_y,k_z) U_{\cal R}^{\dagger},
\end{align} 
with $U_\mathcal{P} = \tau_2 \sigma_0 \rho_0$, $U_\mathcal{T} = \tau_1 \sigma_2 \rho_0$, and $U_\mathcal{R} = \tau_3 \sigma_1 \rho_0$. We further include a symmetry-allowed hybridization,
\begin{align}
  \delta H = & b_1 \tau_3 \sigma_1 \rho_1 + b_2 \tau_1 \sigma_2 \mu_1 \sin k_z \nonumber \\
  &\, \mbox{} + b_3 \tau_2 \sigma_2 \mu_3 \sin k_z + b_4 \tau_2 \sigma_2 \mu_0 \sin k_z , 
\end{align}
with $b_1 = 0.2t$, $b_2 = b_3 = 0.4t$, and $b_4 = 0.6t$, to remove accidental degeneracies and boundary states. The bulk and rhombic pillar band structure as well as the nodal manifold are shown in Fig. ~\ref{fig:app_bandstructures}. Around $k_z = 0$, the system hosts a zero-energy hinge arc. These states hybridize and disappear at a Dirac node of the surface band structure. Since this model corresponds to the element ``2'' in the classification sequence for generic $0 < |k_z| < \pi$, as a function of $k_z$ the transition from the second-order phase at $k_z = 0$ to the trivial phase at $k_z = \pi$ involves two gap closings, separated by an intermediate second-order phase. As shown in Fig.\ \ref{fig:app_bandstructures}, between the first and the second nodal loop, a non-degenerate arc of zero-energy hinge states appears, which corresponds to the second-order topological phase for the value "1" of the crystalline topological invariant at finite $k_z$. 

\section{Classification} 
\label{app:a}

\begin{table*}
\begin{tabular}{ccccc|cc|cc|c}
 &  &  &  &  & \multicolumn{2}{c|}{$k_{z}=0,\pi$} & \multicolumn{2}{c|}{$0<|k_{z}|<\pi$} & \tabularnewline
Class & $\mathcal{T}^{2}$ & $\mathcal{P}^{2}$ & $\mathcal{C}^{2}$ & $\mathcal{S}$ & Class & $\mathcal{K}^{\prime\prime}\subseteq\mathcal{K^{\prime}\subseteq\mathcal{K}}$ & Class & $\mathcal{K}^{\prime\prime}\subseteq\mathcal{K^{\prime}\subseteq\mathcal{K}}$ & type \tabularnewline
\hline 
$\text{AIII}$ & 0 & 0 & 1 & $\mathcal{M}_{+}^{x/y}$ & $\text{AIII}^{\mathcal{M}_{+}}$ & $0\subseteq\mathbb{Z}\subseteq\mathbb{Z}$ & $\text{AIII}^{\mathcal{M}_{+}}$ & $0\subseteq\mathbb{Z}\subseteq\mathbb{Z}$ & (i) \tabularnewline
$\text{A}$ & 0 & 0 & 0 & $\mathcal{CM}^{x/y}$ & $\text{A}^{\mathcal{CM}}$ & $0\subseteq\mathbb{Z}\subseteq\mathbb{Z}^{2}$ & $\text{A}^{\mathcal{CM}}$ & $0\subseteq\mathbb{Z}\subseteq\mathbb{Z}^{2}$ & (i) \tabularnewline
\hline 
BDI & 1 & 1 & 1 & $\mathcal{M}_{++}^{x/y}$ & $\text{BDI}^{\mathcal{M}_{++}}$ & $0\subseteq\mathbb{Z}\subseteq\mathbb{Z}$ & $\text{AIII}^{\mathcal{M}_{+}}$ & $0\subseteq\mathbb{Z}\subseteq\mathbb{Z}$ & (i) \tabularnewline
DIII & -1 & 1 & 1 & $\mathcal{M}_{++}^{x/y}$ & $\text{DIII}^{\mathcal{M}_{++}}$ & $0\subseteq\mathbb{Z}_{2}\subseteq\mathbb{Z}_{2}$ & $\text{AIII}^{\mathcal{M}_{+}}$ & $0\subseteq\mathbb{Z}\subseteq\mathbb{Z}$ & (iv)\tabularnewline
CII & -1 & -1 & 1 & $\mathcal{M}_{++}^{x/y}$ & $\text{CII}^{\mathcal{M}_{++}}$ & $0\subseteq2\mathbb{Z}\subseteq2\mathbb{Z}$ & $\text{AIII}^{\mathcal{M}_{+}}$ & $0\subseteq\mathbb{Z}\subseteq\mathbb{Z}$ & (iv)/(i) \tabularnewline
CI & 1 & -1 & 1 & $\mathcal{M}_{++}^{x/y}$ & $\text{CI}^{\mathcal{M}_{++}}$ & $0\subseteq0\subseteq0$ & $\text{AIII}^{\mathcal{M}_{+}}$ & $0\subseteq\mathbb{Z}\subseteq\mathbb{Z}$ & (iv)\tabularnewline
\hline 
AI & 1 & 0 & 0 & $\mathcal{CM}_{-}^{x/y}$ & $\text{AI}^{\mathcal{CM}_{-}}$ & $0\subseteq0\subseteq0$ & $\text{A}^{\mathcal{CM}}$ & $0\subseteq\mathbb{Z}\subseteq\mathbb{Z}^{2}$ & (iv)\tabularnewline
D & 0 & 1 & 0 & $\mathcal{CM}_{+}^{x/y}$ & $\text{D}^{\mathcal{CM}_{+}}$ & $0\subseteq\mathbb{Z}\subseteq\mathbb{Z}^{2}$ & $\text{A}^{\mathcal{CM}}$ & $0\subseteq\mathbb{Z}\subseteq\mathbb{Z}^{2}$ & (i) \tabularnewline
AII & -1 & 0 & 0 & $\mathcal{CM}_{-}^{x/y}$ & $\text{AII}^{\mathcal{CM}_{-}}$ & $0\subseteq\mathbb{Z}_{2}\subseteq\mathbb{Z}_{2}^{2}$ & $\text{A}^{\mathcal{CM}}$ & $0\subseteq\mathbb{Z}\subseteq\mathbb{Z}^{2}$ & (iv)\tabularnewline
C & 0 & -1 & 0 & $\mathcal{CM}_{+}^{x/y}$ & $\text{C}^{\mathcal{CM}_{+}}$ & $0\subseteq2\mathbb{Z}\subseteq2\mathbb{Z}^{2}$ & $\text{A}^{\mathcal{CM}}$ & $0\subseteq\mathbb{Z}\subseteq\mathbb{Z}^{2}$ & (iv)/(i) \tabularnewline
\hline 
BDI & 1 & 1 & 1 & $\mathcal{M}_{--}^{x/y}$ & $\text{BDI}^{\mathcal{M}_{--}}$ & $0\subseteq0\subseteq0$ & $\text{AIII}^{\mathcal{M}_{+}}$ & $0\subseteq\mathbb{Z}\subseteq\mathbb{Z}$ & (iv)\tabularnewline
DIII & -1 & 1 & 1 & $\mathcal{M}_{--}^{x/y}$ & $\text{DIII}^{\mathcal{M}_{--}}$ & $0\subseteq2\mathbb{Z}\subseteq\mathbb{Z}$ & $\text{AIII}^{\mathcal{M}_{+}}$ & $0\subseteq\mathbb{Z}\subseteq\mathbb{Z}$ & (ii)/(i) \tabularnewline
CII & -1 & -1 & 1 & $\mathcal{M}_{--}^{x/y}$ & $\text{CII}^{\mathcal{M}_{--}}$ & $0\subseteq\mathbb{Z}_{2}\subseteq\mathbb{Z}_{2}$ & $\text{AIII}^{\mathcal{M}_{+}}$ & $0\subseteq\mathbb{Z}\subseteq\mathbb{Z}$ & (iv)\tabularnewline
CI & 1 & -1 & 1 & $\mathcal{M}_{--}^{x/y}$ & $\text{CI}^{\mathcal{M}_{--}}$ & $0\subseteq2\mathbb{Z}\subseteq2\mathbb{Z}$ & $\text{AIII}^{\mathcal{M}_{+}}$ & $0\subseteq\mathbb{Z}\subseteq\mathbb{Z}$ & (iv)/(i) \tabularnewline
\hline 
AI & 1 & 0 & 0 & $\mathcal{CM}_{+}^{x/y}$ & $\text{AI}^{\mathcal{CM}_{+}}$ & $0\subseteq2\mathbb{Z}\subseteq2\mathbb{Z}$ & $\text{A}^{\mathcal{CM}}$ & $0\subseteq\mathbb{Z}\subseteq\mathbb{Z}^{2}$ & (iv)/(i) \tabularnewline
D & 0 & 1 & 0 & $\mathcal{CM}_{-}^{x/y}$ & $\text{D}^{\mathcal{CM}_{-}}$ & $0\subseteq0\subseteq2\mathbb{Z}$ & $\text{A}^{\mathcal{CM}}$ & $0\subseteq\mathbb{Z}\subseteq\mathbb{Z}^{2}$ & (iv)\tabularnewline
AII & -1 & 0 & 0 & $\mathcal{CM}_{+}^{x/y}$ & $\text{AII}^{\mathcal{CM}_{+}}$ & $0\subseteq2\mathbb{Z}\subseteq\mathbb{Z}$ & $\text{A}^{\mathcal{CM}}$ & $0\subseteq\mathbb{Z}\subseteq\mathbb{Z}^{2}$ & (ii)/(i) \tabularnewline
C & 0 & -1 & 0 & $\mathcal{CM}_{-}^{x/y}$ & $\text{C}^{\mathcal{CM}_{-}}$ & $0\subseteq0\subseteq2\mathbb{Z}$ & $\text{A}^{\mathcal{CM}}$ & $0\subseteq\mathbb{Z}\subseteq\mathbb{Z}^{2}$ & (iv)\tabularnewline
\end{tabular}
\caption{
Combination of tenfold-way symmetries ${\cal T}$, ${\cal P}$, or ${\cal C} = {\cal P}{\cal T}$ and mirror symmetry $\mathcal{S}=\mathcal{M}^{x/y}$ that allow for a second-order semimetal or nodal superconductor with anomalous hinge states around $k_z = 0$ (of type (i), (ii), or (iii), as discussed in the text) and/or with flat hinge arcs that occur only between nodal manifolds at finite $k_z$ and do not cross $k_z = 0, \pi$ (type (iv)). The table also contains the symmetry class of the two-dimensional Hamiltonian $H_{k_z}(k_x, k_y)$ at $k_z = 0$, $\pi$ and at $0 < |k_z| < \pi$, together with the corresponding subgroup sequence $\mathcal{K}^{\prime \prime} \subseteq \mathcal{K}^\prime \subseteq \mathcal{K}$. The rightmost column displays the type of boundary phenomenology around $k_z = 0$ that is induced by a nontrivial higher-order topology of $H_{k_z}(k_x,k_y)$ for generic $0 < |k_z| < \pi$. If multiple types are listed, the first and second entries are for odd and even multiples of the generator of the nontrivial topology, respectively. For the types ``(i)'', ``(ii)'', and ``(iii)'', it is assumed that $H_{k_z}(k_x,k_y)$ does not go through a topological phase transition if $k_z \to 0$, whereas ``(iv)'' refers to a topological phase of $H_{k_z}(k_x,k_y)$ at generic $k_z$ that does not come with higher-order boundary signatures upon going to $k_z = 0$ or that is incompatible with the symmetry constraints at $k_z =0$, so that a gap closing is necessary upon going to $k_z = 0$. 
\label{tab:class_Mxy}}
\end{table*}

\begin{table*}
\begin{tabular}{ccccc|cc|cc|c}
 &  &  &  &  & \multicolumn{2}{c|}{$k_{z}=0,\pi$} & \multicolumn{2}{c|}{$0<|k_{z}|<\pi$} & \tabularnewline
Class & $\mathcal{T}^{2}$ & $\mathcal{P}^{2}$ & $\mathcal{C}^{2}$ & $\mathcal{S}$ & Class & $\mathcal{K}^{\prime\prime}\subseteq\mathcal{K^{\prime}\subseteq\mathcal{K}}$ & Class & $\mathcal{K}^{\prime\prime}\subseteq\mathcal{K^{\prime}\subseteq\mathcal{K}}$ & type \tabularnewline
\hline 
$\text{AIII}$ & 0 & 0 & 1 & $\mathcal{P}^{+}\mathcal{R}_{+}^{x/y}$ & $\text{AIII}^{\mathcal{P}^{+}\mathcal{M}_{+}}$ & $0\subseteq\mathbb{Z}_{2}\subseteq\mathbb{Z}_{2}$ & $\text{AIII}^{\mathcal{P}^{+}\mathcal{M}_{+}}$ & $0\subseteq\mathbb{Z}_{2}\subseteq\mathbb{Z}_{2}$ & (i)\tabularnewline
$\text{A}$ & 0 & 0 & 0 & $\mathcal{P}^{+}\mathcal{R}^{x/y}$ & $\text{A}^{\mathcal{P}^{+}\mathcal{M}}$ & $0\subseteq\mathbb{Z}_{2}\subseteq\mathbb{Z}_{2}$ & $\text{A}^{\mathcal{P}^{+}\mathcal{M}}$ & $0\subseteq\mathbb{Z}_{2}\subseteq\mathbb{Z}_{2}$ & (i)\tabularnewline
\hline 
BDI & 1 & 1 & 1 & $\mathcal{R}_{++}^{x/y}$ & $\text{BDI}^{\mathcal{M}_{++}}$ & $0\subseteq\mathbb{Z}\subseteq\mathbb{Z}$ & $\text{AIII}^{\mathcal{P}^{+}\mathcal{M}_{+}}$ & $0\subseteq\mathbb{Z}_{2}\subseteq\mathbb{Z}_{2}$ & (i)\tabularnewline
D & 0 & 1 & 0 & $\mathcal{R}_{+}^{x/y}$ & $\text{D}^{\mathcal{M}_{+}}$ & $0\subseteq\mathbb{Z}_{2}\subseteq\mathbb{Z}_{2}$ & $\text{A}^{\mathcal{P}^{+}\mathcal{M}}$ & $0\subseteq\mathbb{Z}_{2}\subseteq\mathbb{Z}_{2}$ & (i)\tabularnewline
\hline 
DIII & -1 & 1 & 1 & $\mathcal{R}_{-+}^{x/y}$ & $\text{DIII}^{\mathcal{M}_{-+}}$ & $0\subseteq\mathbb{Z}_{2}\subseteq\mathbb{Z}_{2}^{2}$ & $\text{AIII}^{\mathcal{P}^{+}\mathcal{M}_{+}}$ & $0\subseteq\mathbb{Z}_{2}\subseteq\mathbb{Z}_{2}$ & (ii)\tabularnewline
AII & -1 & 0 & 0 & $\mathcal{CR}_{-}^{x/y}$ & $\text{AII}^{\mathcal{CM}_{-}}$ & $0\subseteq\mathbb{Z}_{2}\subseteq\mathbb{Z}_{2}^{2}$ & $\text{A}^{\mathcal{P}^{+}\mathcal{M}}$ & $0\subseteq\mathbb{Z}_{2}\subseteq\mathbb{Z}_{2}$ & (ii)\tabularnewline
\hline 
CII & -1 & -1 & 1 & $\mathcal{R}_{--}^{x/y}$ & $\text{CII}^{\mathcal{M}_{--}}$ & $0\subseteq\mathbb{Z}_{2}\subseteq\mathbb{Z}_{2}$ & $\text{AIII}^{\mathcal{P}^{+}\mathcal{M}_{+}}$ & $0\subseteq\mathbb{Z}_{2}\subseteq\mathbb{Z}_{2}$ & (iv)\tabularnewline
C & 0 & -1 & 0 & $\mathcal{R}_{-}^{x/y}$ & $\text{C}^{\mathcal{M}_{-}}$ & $0\subseteq0\subseteq0$ & $\text{A}^{\mathcal{P}^{+}\mathcal{M}}$ & $0\subseteq\mathbb{Z}_{2}\subseteq\mathbb{Z}_{2}$ & (iv)\tabularnewline
\hline 
CI & 1 & -1 & 1 & $\mathcal{R}_{+-}^{x/y}$ & $\text{CI}^{\mathcal{M}_{+-}}$ & $0\subseteq0\subseteq0$ & $\text{AIII}^{\mathcal{P}^{+}\mathcal{M}_{+}}$ & $0\subseteq\mathbb{Z}_{2}\subseteq\mathbb{Z}_{2}$ & (iv)\tabularnewline
AI & 1 & 0 & 0 & $\mathcal{CR}_{+}^{x/y}$ & $\text{AI}^{\mathcal{CM}_{+}}$ & $0\subseteq\mathbb{Z}\subseteq\mathbb{Z}$ & $\text{A}^{\mathcal{P}^{+}\mathcal{M}}$ & $0\subseteq\mathbb{Z}_{2}\subseteq\mathbb{Z}_{2}$ & (i)\tabularnewline
\end{tabular}
\caption{ 
Same as table \ref{tab:class_Mxy}, but for rotation symmetry $\mathcal{S}=\mathcal{R}^{x/y}$.
\label{tab:class_Rxy}}
\end{table*}

\begin{table*}
\begin{tabular}{ccccc|cc|cc|c}
 &  &  &  &  & \multicolumn{2}{c|}{$k_{z}=0,\pi$} & \multicolumn{2}{c|}{$0<|k_{z}|<\pi$} & \tabularnewline
Class & $\mathcal{T}^{2}$ & $\mathcal{P}^{2}$ & $\mathcal{C}^{2}$ & $\mathcal{S}$ & Class & $\mathcal{K}^{\prime\prime}\subseteq\mathcal{K^{\prime}\subseteq\mathcal{K}}$ & Class & $\mathcal{K}^{\prime\prime}\subseteq\mathcal{K^{\prime}\subseteq\mathcal{K}}$ & type\tabularnewline
\hline 
$\text{AIII}$ & 0 & 0 & 1 & $\mathcal{R}_{-}^{z}$ & $\text{AIII}^{\mathcal{R}_{-}}$ & $2\mathbb{Z}\subseteq\mathbb{Z}\subseteq\mathbb{Z}$ & $\text{AIII}^{\mathcal{R}_{-}}$ & $2\mathbb{Z}\subseteq\mathbb{Z}\subseteq\mathbb{Z}$ & (i)/(iv) \tabularnewline
\hline 
BDI & 1 & 1 & 1 & $\mathcal{R}_{+-}^{z}$ & $\text{BDI}^{\mathcal{R}_{+-}}$ & $2\mathbb{Z}\subseteq\mathbb{Z}\subseteq\mathbb{Z}$ & $\text{AIII}^{\mathcal{R}_{-}}$ & $2\mathbb{Z}\subseteq\mathbb{Z}\subseteq\mathbb{Z}$ & (i)/(iv) \tabularnewline
DIII & -1 & 1 & 1 & $\mathcal{R}_{-+}^{z}$ & $\text{DIII}^{\mathcal{R}_{-+}}$ & $0\subseteq\mathbb{Z}_{2}\subseteq\mathbb{Z}_{2}$ & $\text{AIII}^{\mathcal{R}_{-}}$ & $2\mathbb{Z}\subseteq\mathbb{Z}\subseteq\mathbb{Z}$ & (iv)\tabularnewline
CII & -1 & -1 & 1 & $\mathcal{R}_{+-}^{z}$ & $\text{CII}^{\mathcal{R}_{+-}}$ & $4\mathbb{Z}\subseteq2\mathbb{Z}\subseteq2\mathbb{Z}$ & $\text{AIII}^{\mathcal{R}_{-}}$ & $2\mathbb{Z}\subseteq\mathbb{Z}\subseteq\mathbb{Z}$ & (iv)/(i){*}/(iv)/(iv)\tabularnewline
CI & 1 & -1 & 1 & $\mathcal{R}_{-+}^{z}$ & $\text{CI}^{\mathcal{R}_{-+}}$ & $0\subseteq0\subseteq0$ & $\text{AIII}^{\mathcal{R}_{-}}$ & $2\mathbb{Z}\subseteq\mathbb{Z}\subseteq\mathbb{Z}$ & (iv)\tabularnewline
\hline 
D & 0 & 1 & 0 & $\mathcal{R}_{-}^{z}$ & $\text{D}^{\mathcal{R}_{-}}$ & $2\mathbb{Z}\subseteq\mathbb{Z}\subseteq\mathbb{Z}^{2}$ & $\text{A}^{\mathcal{R}}$ & $\mathbb{Z}\subseteq\mathbb{Z}\subseteq\mathbb{Z}^{2}$ & (iii)/(iv) \tabularnewline
\hline 
BDI & 1 & 1 & 1 & $\mathcal{R}_{-+}^{z}$ & $\text{BDI}^{\mathcal{R}_{-+}}$ & $0\subseteq0\subseteq0$ & $\text{AIII}^{\mathcal{R}_{-}}$ & $2\mathbb{Z}\subseteq\mathbb{Z}\subseteq\mathbb{Z}$ & (iv)\tabularnewline
DIII & -1 & 1 & 1 & $\mathcal{R}_{+-}^{z}$ & $\text{DIII}^{\mathcal{R}_{+-}}$ & $4\mathbb{Z}\subseteq2\mathbb{Z}\subseteq\mathbb{Z}$ & $\text{AIII}^{\mathcal{R}_{-}}$ & $2\mathbb{Z}\subseteq\mathbb{Z}\subseteq\mathbb{Z}$ & (ii)/(iii)/(ii)/(iv) \tabularnewline
CII & -1 & -1 & 1 & $\mathcal{R}_{-+}^{z}$ & $\text{CII}^{\mathcal{R}_{-+}}$ & $0\subseteq\mathbb{Z}_{2}\subseteq\mathbb{Z}_{2}$ & $\text{AIII}^{\mathcal{R}_{-}}$ & $2\mathbb{Z}\subseteq\mathbb{Z}\subseteq\mathbb{Z}$ & (iv)\tabularnewline
CI & 1 & -1 & 1 & $\mathcal{R}_{+-}^{z}$ & $\text{CI}^{\mathcal{R}_{+-}}$ & $2\mathbb{Z}\subseteq2\mathbb{Z}\subseteq2\mathbb{Z}$ & $\text{AIII}^{\mathcal{R}_{-}}$ & $2\mathbb{Z}\subseteq\mathbb{Z}\subseteq\mathbb{Z}$ & (iv)\tabularnewline
\end{tabular}
\caption{
Same as table \ref{tab:class_Mxy}, but for rotation symmetry $\mathcal{S}=\mathcal{R}^{z}$. If four types are listed in the rightmost column, the $n$th element refers to the boundary phenomenology for the case that the topological equivalence class of $H_{k_z}(k_x,k_y)$ at generic $k_z$ is that of the ($4m + n$)-fold multiple of the generator of the nontrivial topology, with $m$ integer. The type (i)$^*$ is discussed in detail in the text, see the discussion preceding Eq. \eqref{eq:H_CII_Rz+-}.
\label{tab:class_Rz}}
\end{table*}

\begin{table*}
\begin{tabular}{ccccc|cc|cc|c}
 &  &  &  &  & \multicolumn{2}{c|}{$k_{z}=0,\pi$} & \multicolumn{2}{c|}{$0<|k_{z}|<\pi$} & \tabularnewline
Class & $\mathcal{T}^{2}$ & $\mathcal{P}^{2}$ & $\mathcal{C}^{2}$ & $\mathcal{S}$ & Class & $\mathcal{K}^{\prime\prime}\subseteq\mathcal{K^{\prime}\subseteq\mathcal{K}}$ & Class & $\mathcal{K}^{\prime\prime}\subseteq\mathcal{K^{\prime}\subseteq\mathcal{K}}$ & type \tabularnewline
\hline 
$\text{AIII}$ & 0 & 0 & 1 & $\mathcal{T}^{+}\mathcal{I}_{-}$ & $\text{AIII}^{\mathcal{T}^{+}\mathcal{R}_{-}}$ & $0\subseteq\mathbb{Z}_{2}\subseteq\mathbb{Z}_{2}$ & $\text{AIII}^{\mathcal{T}^{+}\mathcal{R}_{-}}$ & $0\subseteq\mathbb{Z}_{2}\subseteq\mathbb{Z}_{2}$ & (i)\tabularnewline
\hline 
CI & 1 & -1 & 1 & $\mathcal{I}_{++}$ & $\text{CI}^{\mathcal{R}_{++}}$ & $0\subseteq0\subseteq0$ & $\text{AIII}^{\mathcal{T}^{+}\mathcal{R}_{-}}$ & $0\subseteq\mathbb{Z}_{2}\subseteq\mathbb{Z}_{2}$ & (iv)\tabularnewline
\hline 
BDI & 1 & 1 & 1 & $\mathcal{I}_{+-}$ & $\text{BDI}^{\mathcal{R}_{+-}}$ & $2\mathbb{Z}\subseteq\mathbb{Z}\subseteq\mathbb{Z}$ & $\text{AIII}^{\mathcal{T}^{+}\mathcal{R}_{-}}$ & $0\subseteq\mathbb{Z}_{2}\subseteq\mathbb{Z}_{2}$ & (i) \tabularnewline
D & 0 & 1 & 0 & $\mathcal{CI}_{+}$ & $\text{D}^{\mathcal{CR}_{+}}$ & $0\subseteq\mathbb{Z}_{2}\subseteq\mathbb{Z}_{2}$ & $\text{A}^{\mathcal{T}^{+}\mathcal{R}}$ & $\mathbb{Z}_{2}\subseteq\mathbb{Z}_{2}\subseteq\mathbb{Z}_{2}$ & (iii) \tabularnewline
\hline 
DIII & -1 & 1 & 1 & $\mathcal{I}_{--}$ & $\text{DIII}^{\mathcal{R}_{--}}$ & $0\subseteq\mathbb{Z}_{2}\subseteq\mathbb{Z}_{2}^{2}$ & $\text{AIII}^{\mathcal{T}^{+}\mathcal{R}_{-}}$ & $0\subseteq\mathbb{Z}_{2}\subseteq\mathbb{Z}_{2}$ & (ii) \tabularnewline
\hline 
CII & -1 & -1 & -1 & $\mathcal{I}_{-+}$ & $\text{CII}^{\mathcal{R}_{-+}}$ & $0\subseteq\mathbb{Z}_{2}\subseteq\mathbb{Z}_{2}$ & $\text{AIII}^{\mathcal{T}^{+}\mathcal{R}_{-}}$ & $0\subseteq\mathbb{Z}_{2}\subseteq\mathbb{Z}_{2}$ & (iv)\tabularnewline
\end{tabular}
\caption{
Same as table \ref{tab:class_Mxy}, but for inversion symmetry $\mathcal{S}=\mathcal{I}$.
\label{tab:class_I}}
\end{table*}

Tables \ref{tab:class_Mxy}--\ref{tab:class_I} contain a complete list of all combinations of tenfold-way symmetries and an additional order-two crystalline symmetry, for which second-order gapless topological phases are possible. The crystalline symmetries are mirror ${\cal M}^{x/y}$, twofold rotation ${\cal R}^{x/y}$ around the $x$ or $y$ axis, twofold rotation ${\cal R}^z$ around the $z$ axis, and inversion ${\cal I}$. The tables also contain unitary and antiunitary crystalline antisymmetries, which are denoted as the product of the chiral conjugation operation ${\cal C}$ or particle-hole conjugation ${\cal P}$ (squaring to one) and a crystalline symmetry. The subscripts $\pm$ indicate, whether the crystalline symmetry ${\cal S}$ commutes ($+$) or anticommutes $(-)$ with ${\cal T}$, ${\cal P}$, or ${\cal C}$. (If there are two subscripts, the first subscript describes the commutation or anticommutation with ${\cal T}$, whereas the second subscript refers to ${\cal P}$.)
The tables also contain the crystalline symmetry classes of the two-dimensional Hamiltonian $H_{k_z}(k_x,k_y)$ at the high-symmetry momenta $k_z = 0$, $\pi$ and at generic $0 < |k_z| < \pi$, together with the corresponding subgroup sequences $\mathcal{K}^{\prime \prime} \subseteq \mathcal{K}^\prime \subseteq \mathcal{K}$.

Within the same symmetry class, the second-order boundary phenomenology may be different for the generators of the nontrivial higher-order topology and for direct sums of these generators. An example is the tenfold-way symmetry class DIII with a mirror symmetry ${\cal M}^x_{--}$, see Eqs.\ (\ref{eq:H_DIII_M--}) and (\ref{eq:H_DIII_M--_type1}), for which the generator of the nontrivial topology has boundary states of type (ii) and the direct sum of the generator with itself has boundary states of type (i). The rightmost columns of Tables \ref{tab:class_Mxy}--\ref{tab:class_I} list other examples of symmetry classes, for which taking the direct sums of the generator of the nontrivial topology leads to a different boundary phenomenology. To be precise, for the right columns of Tables \ref{tab:class_Mxy}--\ref{tab:class_I} we consider nodal semimetals or superconductors obtained by taking multiple direct sums of the generator of the nontrivial higher-order topology of $H_{k_z}(k_x,k_y)$ at a reference value $k_z = k_{z,{\rm ref}}$, with $0 < |k_{z,{\rm ref}}| < \pi$. If $H_{k_z}(k_x,k_y)$ can be continuously ({\em i.e.}, without gap closing) extended to $k_z = 0$, the three-dimensional Hamiltonian $H(k_x,k_y,k_z)$ corresponding to $H_{k_z}(k_x,k_y)$ has a unique nontrivial boundary phenomenology at $k_z = 0$, which is one of the types (i) or (ii) discussed in the main text. If a continuous extension to $k_z = 0$ is possible, but there is no nontrivial boundary signature at $k_z = 0$, or if a continuous extension to $k_z = 0$ without gap closing is not possible, we use the label ``(iv)'' to indicate that higher-order boundary signatures occur away from $k_z = 0$. Additionally, right column of Tables \ref{tab:class_Rz} and \ref{tab:class_I} contain two symmetry classes with boundary phenomenology of type (iii), which originates from the generators of the nontrivial topology at $k_z = 0$, while $H_{k_z}(k_x,k_y)$ is an obstructed atomic limit at generic $0 < |k_z| < \pi$.

To identify all second-order topological semimetals and nodal superconductors and their types as summarized in tables \ref{tab:class_Mxy}--\ref{tab:class_I}, we made use of the order-resolved classification of two-dimensional second-order topological insulators and superconductors from Refs.~\onlinecite{geier2018, trifunovic2019} and performed a few additional checks as follows. First, a necessary and sufficient criterion of a second-order gapless topological phase of type (i), (ii), or (iv) is a two-dimensional second-order topological phase at $0 < |k_z| < \pi$. All symmetry classes with an order-two symmetry fulfilling this criteria are included in the tables \ref{tab:class_Mxy}--\ref{tab:class_I}. To identify the type, one needs to compute the forgetful functor from the classifying group at $k_z = 0$, $\pi$ to the classifying group at finite $0 < |k_z| < \pi$, which can be done either by inspecting the generator Hamiltonians and/or by inspecting the boundary signatures of the corresponding topological phases, \textit{e.g.} with the help of Ref.~\onlinecite{geier2018}. Second-order gapless phases of type (iii) (with chiral or helical hinge states) may appear when the Hamiltonian at $k_z = 0$ or $\pi$ is in a second-order topological phase, which becomes an obstructed atomic limit at $0<|k_z|<\pi$ under the forgetful functor. The positive results for this type are included in the tables \ref{tab:class_Mxy}--\ref{tab:class_I}.

Table \ref{tab:classification} in the main text contains the possible types of higher-order topological semimetals and nodal superconductors with a unitary or antiunitary order-two crystalline symmetry as extracted from tables \ref{tab:class_Mxy}--\ref{tab:class_I}. While in tables \ref{tab:class_Mxy}--\ref{tab:class_I} we followed the convention of Refs. \onlinecite{geier2018, trifunovic2019} for the labeling of the real crystalline symmetry classes with a unitary symmetry $\mathcal{S}$ or antisymmetry $\mathcal{CS}$ for mathematical simplicity, in table \ref{tab:classification} in the main text we preferred the more physically relevant notation using unitary symmetries $\mathcal{S}$ or antiunitary symmetries $\mathcal{T}^\pm\mathcal{S}$. The relation between the two schemes follows by combining the crystalline symmetry operators with the tenfold-way symmetries, \textit{i.e.} in class D (C) we can label an additional order-two crystalline symmetry either by $\mathcal{CS}_\pm$ or equivalently by $\mathcal{P}\mathcal{CS}_\pm = \mathcal{T}^\pm \mathcal{S}_\pm$ ($\mathcal{T}^\mp \mathcal{S}_\pm$), and in class AIII we have $\mathcal{C} \mathcal{P}^s \mathcal{S}_\pm = \mathcal{T}^{\pm s} \mathcal{S}_\pm$ with $s = \pm 1$.

\section{Disorder}
\label{app:dis}

Disorder breaks translation symmetry and any crystalline symmetries that protect the hinge states. However, if the bulk density of states at the nodal points is sufficiently small, which requires the absence of trivial bulk bands or type-II Weyl nodes, the hybridization of hinge states with bulk states is strongly suppressed for weak disorder. In that case, the dominant effect of disorder is a hybridization between hinge states, leading to a disorder-broadened arc of localized hinge states.

As discussed in the main text, for a disorder potential that obeys the tenfold-way symmetries, such disorder-induced hybridization of hinge states can be forbidden by the symmetry constraints.
Whether or not this is the case can be read off from the tables for the ``extrinsic'' classification group $\bar{\cal K}_{\rm e}$ of anomalous corner states of two-dimensional second-order topological phases in the presence of disorder \cite{geier2018}. In Ref.\ \onlinecite{geier2018}, the label ``extrinsic'' indicates that corner states may arise as a result of a nontrivial bulk topology or as a result of a symmetry-compatible ``decoration'' of the crystal boundaries with topologically nontrivial insulators or superconductors. 
Since disorder breaks both the order-two crystalline symmetry as well as the translation symmetry along $z$, hinge disorder ``collapses'' the one-dimensional hinge states into a collection of zero-dimensional corner states. Thus, it is the extrinsic classification group $\bar{\mathcal{K}}_e$ of Ref.\ \onlinecite{geier2018} for the crystalline symmetry group appropriate for $k_z = 0$ that classifies possible hinge states of second-order topological gapless phases in the presence of disorder.
In case the extrinsic classification group $\bar{\mathcal{K}}_e \simeq \ZZ$, zero-energy hinge states all must have the same eigenvalue under a unitary antisymmetry $\mathcal{C}$, which prohibits their hybridization. In case of $\bar{\mathcal{K}}_e \simeq \ZZ_2$, the hinge states are generally allowed to hybridize in pairs, such that only the number parity of (Kramers pairs of) zero-energy states per hinge must remain after hybridization (in the presence of time-reversal symmetry $\mathcal{T}^2 = -1$, if present). In case of $\bar{\mathcal{K}}_e \simeq 0$, the hinge states are not only allowed to hybridize in pairs, but they generically acquire a finite energy themselves. 
In the following we illustrate the consequences of this protection of hinge states for the examples discussed in the main text and this appendix.

{\it Model of Eq. \eqref{eq:H_AIII_My+}. ---} In class AIII, the zero-energy hinge states are eigenstates under the chiral conjugation $\mathcal{C}$. The eigenvalue $c = \pm 1$ assigns a ``chirality'' to the zero-energy states. Any perturbation preserving the chiral antisymmetry cannot hybridize eigenstates with the same chirality eigenvalue. In this crystalline symmetry class, hinge states on the same hinge have the same chirality eigenvalue. Therefore, disorder cannot hybridize hinge-states on the same hinge.

This stability of the edge states is consistent with the extrinsic classification group $\bar{\mathcal{K}}_e \simeq \mathbb{Z}$ of mirror-symmetry breaking corners of two-dimensional gapped systems in class AIII with mirror symmetry $\mathcal{M}_+$, see Ref. \onlinecite{geier2018}.

{\it Models of Eq. \eqref{eq:H_DIII_M--} and Eq.\ \eqref{eq:H_DIII_M--_type1}. ---}
In class DIII, eigenstates come in Kramers pairs, where the two partners in the Kramers pair have opposite chirality under the chiral antisymmetry $\mathcal{C} = \mathcal{PT}$. For the hinge states with finite $k_z$, time-reversal symmetry relates hinge states with opposite $k_z$ and opposite chirality eigenvalues. Although time-reversal symmetry forbids that such Kramers pairs can hybridize and gap out in pairs, nonzero matrix elements may exist between hinge states at values of $k_z$ that differ in sign and magnitude, leading to a (weak) hybridization of hinge states. 

This result is consistent with the extrinsic classification group $\bar{\mathcal{K}}_e \simeq \ZZ_2$ \cite{geier2018}. With broken translation symmetry, all Kramers pairs formed from modes with opposite $k_z$ hybridize and gap out in pairs. After hybridization, a zero-energy hinge state remains if the number of zero-energy Kramers pairs per hinge is odd. The parity of Kramers pairs per hinge in general depends on the geometry and boundary conditions at top and bottom of the hinge. Notice that there is no Kramers pair at $k_z = 0$, since the node of the Dirac cone on the adjacent surfaces lies at $k_z = 0$.

Notice that, because the chirality eigenvalues of hinge states on a single hinge at $k_z > 0$ and $k_z < 0 $ must be opposite, the surface states at $k_z = 0$ must be gapless in order to support the change in boundary topology between $k_z > 0$ and $k_z < 0 $. This change of boundary topology manifests itself as the first-order surface states at $k_z = 0$ in this higher-order topological gapless phase of type (ii). 

{\it Model Eq. \eqref{eq:H_D_Ry+}. ---} For this crystalline symmetry class, hinge states are protected by a $\ZZ_2$ invariant, so that hybridization between hinge states at different $k_z$ is always possible. The parity of zero-energy modes per hinge, and thus the question whether a zero-energy hinge mode remains after hybridization, depends on the sample geometry and boundary conditions. This is in agreement with the extrinsic classifying group $\bar{\mathcal{K}}_e \simeq \ZZ_2$ for this symmetry class \cite{geier2018}.

{\it Models of Eq. \eqref{eq:H_D_Rz-} and Eq. \eqref{eq:H_DIII_Rz+-}. ---} These examples have chiral or helical hinge states on each hinge. There is only a single zero-energy mode on each hinge, at $k_z=0$. Chiral/helical hinge states are protected from localization. The single mode at zero-energy is protected from gapping by the local symmetries, consistent with the boundary classifying group $\mathcal{K} \simeq \ZZ_2$ with rotation symmetry, {\it c.f.} Table I in Ref.~\onlinecite{geier2018}. Because chiral/helical hinge modes have nonzero energy at $k_z \neq 0$, there is a stronger hybridization with bulk states than for hinges with flat arcs of zero-energy states away from $k_z = 0$. 

{\it Model of Eq. \eqref{eq:H_DIII_Rz-+}/ and Eq.\ \eqref{eq:H_DIII_Rz-+_8bands}.---} For this model hinge states are eigenstates of the chiral conjugation ${\cal C}$. Time-reversal enforces that modes at $k_z$ and $-k_z$ have opposite chirality. As discussed above, time-reversal symmetry protects only modes with opposite $k_z$ from hybridization. The parity of zero-energy Kramers pairs per hinge, and thus whether one must remain after hybridization, depends on the geometry. This results is in agreement with the corresponding boundary classifying group $\mathcal{K} \simeq \ZZ_2$, {\it c.f.} Table I in Ref.~\onlinecite{geier2018}.

{\it Model Eq. \eqref{eq:H_CII_Rz+-}. ---} In this symmetry class, the hinge states on a single hinge have the same eigenvalue under chiral conjugation $\mathcal{C}$. This prohibits their hybridization from hinge disorder. This results is in agreement with the corresponding boundary classifying group $\mathcal{K} \simeq \ZZ$, {\it c.f.} Table I in Ref.~\onlinecite{geier2018}.

\section{Boundary features that do not occur in the presence of a single twofold crystalline symmetry}
\label{app:e}
{\it Symmetry-enforced degeneracy of flat hinge arcs for $k_z \neq 0$. ---} The flat zero-energy hinge arcs of the examples in the main text are non-degenerate. The appendix discusses an example in which a twofold degeneracy is protected by time-reversal symmetry at $k_z = 0$, but the degeneracy is not symmetry-enforced for nonzero $k_z$. In fact, with only a single twofold crystalline symmetry, it is impossible to achieve a symmetry-enforced degeneracy away from the high-symmetry momenta $k_z = 0$, $\pi$. To see this, note that the antiunitary symmetries ${\cal P}$ and ${\cal T}$ that could enforce such a degeneracy change the sign of $k_z$, so that they would have to be combined with a mirror reflection ${\cal M}^z$ to obtain a symmetry constraint of the two-dimensional Hamiltonian $H_{k_z}(k_x,k_y)$ at generic $0 < |k_z| < \pi$ that acts within a hinge. That means that, if only a single crystalline symmetry is present, it must be ${\cal M}^z$. However, if the crystalline symmetry is ${\cal M}^z$, all symmetries of the effective two-dimensional Hamiltonian $H_{k_z}(k_x,k_y)$ are local and only first-order topological phases are possible.

Notice, however, that a second-order topological nodal superconductor with flat hinge arcs with symmetry-enforced degeneracy can be achieved in the presence of two order-two crystalline symmetries. An example of such a second-order phase exists in tenfold-way class DIII with the (unusual) combination of two mirror symmetries ${\cal M}^z$ and ${\cal M}^x$ that both square to one, mutually commute, and commute with $\mathcal{T}$ and $\mathcal{P}$. (Alternatively, ${\cal M}^z$ and ${\cal M}^x$ square to minus one, commute mutually, and anticommute with ${\cal T}$ and ${\cal P}$.) Then, the two-dimensional Hamiltonian $H_{k_z}(k_x,k_y)$ at generic $0 < |k_z| < \pi$ satisfies symmetry constraints corresponding to ${\cal M}^z {\cal P}$, ${\cal M}^z {\cal T}$, and ${\cal M}^x$. This corresponds to the two-dimensional crystalline symmetry class $\mbox{DIII}^{{\cal M}_{++}}$, which admits a second-order phase with Majorana Kramers pairs as corner states \cite{geier2018,trifunovic2019}. Returning to the three-dimensional model, one obtains a second-order topological nodal superconductor with symmetry-enforced twofold degenerate zero-energy hinge Fermi arcs on ${\cal M}^x$-symmetric hinges. A full classification of such topological gapless phases with multiple order-two crystalline symmetries is beyond the scope of this article.

{\it Second-order gapless phases with counter-propagating chiral hinge modes. --- }
The examples of type (iii) second-order topological semimetals or nodal superconductors in Cartan classes D or CII that were discussed in the main text and in the appendix have chiral hinge modes on rotation- or inversion-related hinges that propagate in the same direction. By checking combinations of Cartan classes and order-two crystalline symmetries as shown in Tables \ref{tab:classification} and \ref{tab:class_Mxy}--\ref{tab:class_I}, we find that second-order gapless phases with co-propagating hinge arcs also exist in Cartan class D with rotation symmetry $\mathcal{R}^z_-$ or inversion antisymmetry $\mathcal{CI}_+$. In all cases, both rotation symmetry and inversion antisymmetry require that that the chiral hinge modes on rotation- or inversion-symmetry related hinges must be co-propagating. 

One may ask if there exist type (iii) second-order topological semimetals or nodal superconductors where the crystalline symmetries require the chiral hinge modes at different hinges to be counter-propagating. By checking all symmetry classes, we find that all systems with counter-propagating chiral hinge modes on symmetry related hinges are fully-gapped second order topological phases in three dimensions. Thus, while we do not exclude that type (iii) second-order topological semimetals or nodal superconductors with counter-propagating chiral hinge modes exist for other combinations of crystalline symmetries, they do not exist within the set of tenfold-way classes with a single order-two crystalline symmetry that is considered here.

\section{Crystalline symmetry classes} 
\label{app:d}

The presence of time reversal symmetry ${\cal T}$, particle-hole antisymmetry ${\cal P}$, or their product ${\cal C} = {\cal P}{\cal T}$ and the squares ${\cal T}^2$ and ${\cal P}^2$ (if present) determine the tenfold-way symmetry class of an insulator or superconductor, see Table \ref{tab:classification}. We here consider crystalline symmetry classes, in which one twofold crystalline symmetry ${\cal S}$ is present in addition to the tenfold-way symmetries ${\cal T}$, ${\cal P}$, and/or ${\cal C}$. A symmetry operation is said to be of order two if its square is proportional to the identity operation. In three dimensions, the twofold crystalline symmetries are mirror ${\cal M}$, twofold rotation ${\cal R}$, or inversion ${\cal I}$. The resulting symmetry classes have been enumerated by Shiozaki and Sato \cite{shiozaki2014}. We now briefly explain the construction of the symmetry classes of Ref.\ \onlinecite{shiozaki2014} and the notation of Ref.~\onlinecite{trifunovic2019}, which is used throughout this article.

The crystalline symmetry classes are distinguished by the algebraic relations of their symmetry elements. 
For order-two crystalline symmetries $\mathcal{S}$, it is sufficient to distinguish symmetry operations that square to $+1$ (labeled by $\eta_\mathcal{S} = 1$) and $-1$ ($\eta_\mathcal{S} = -1$).
We introduce the labels $\eta_\mathcal{T}$, $\eta_\mathcal{P}$ and $\eta_\mathcal{C}$ to denote the commutation relations of the crystalline symmetry element $\mathcal{S}$ with the antiunitary symmetry $\mathcal{T}$, antiunitary antisymmetry $\mathcal{P}$ and unitary antisymmetry $\mathcal{C}$.
Unitary crystalline symmetries ($\sigma_\mathcal{S} = 1$) and antisymmetries ($\sigma_\mathcal{S} = -1$) are represented by a unitary matrix $U_\mathcal{S}$ as
\begin{equation}
H(\vk) = \sigma_\mathcal{S} U(\mathcal{S}) H(\mathcal{S} \vk) U_\mathcal{S}^\dagger
\end{equation}
and the representations satisfy the relations
\begin{align*}
U_\mathcal{S}^2 & = \eta_\mathcal{S} \\
U_\mathcal{S} U_\mathcal{T} & = \eta_\mathcal{T} U_\mathcal{T} U^*_\mathcal{S} \\
U_\mathcal{S} U_\mathcal{P} & = \eta_\mathcal{P} U_\mathcal{P} U^*_\mathcal{S} \\
U_\mathcal{S} U_\mathcal{C} & = \eta_\mathcal{C} U_\mathcal{C} U_\mathcal{S} .
\end{align*}
where $\mathcal{S} \vk$ denotes the action of the crystalline symmetry on the momentum coordinates.
Similarly, the representations for antiunitary crystalline symmetries and antisymmetries are defined as
\begin{equation}
H(\vk) = \sigma_\mathcal{S} U_\mathcal{S} H(- \mathcal{S} \vk)^* U_\mathcal{S}^\dagger
\end{equation}
and the representations satisfy the relations
\begin{align*}
U_\mathcal{S} U^*_\mathcal{S} & = \eta_\mathcal{S} \\
U_\mathcal{S} U^*_\mathcal{T} & = \eta_\mathcal{T} U_\mathcal{T} U^*_\mathcal{S} \\
U_\mathcal{S} U^*_\mathcal{P} & = \eta_\mathcal{P} U_\mathcal{P} U^*_\mathcal{S} \\
U_\mathcal{S} U^*_\mathcal{C} & = \eta_\mathcal{C} U_\mathcal{C} U_\mathcal{S} .
\end{align*}

Following this discussion, a tenfold-way class with the additional crystalline symmetry ${\cal S}$ is fully characterized by the tenfold way class, the set of numbers $\eta_{{\cal S}},\eta_{\cal T}, \eta_{\cal P}, \eta_{\cal C}$ and $\sigma_{\cal S}$, and by specifying, whether the crystalline symmetry is unitary or antiunitary \cite{shiozaki2014}. For the purpose of topological classification, this characterization is partially redundant, however.
In particular, 
a Hamiltonian that satisfies a crystalline symmetry $\mathcal{S}$ with representation $U_\mathcal{S}$ also satisfies this crystalline symmetry with representation $i U_\mathcal{S}$. Under this transformation $U_\mathcal{S} \to i U_\mathcal{S}$, the sign of the square of a unitary crystalline symmetry is flipped $\eta_\mathcal{S} \to - \eta_\mathcal{S}$ while at the same time the commutation relation with the antiunitary symmetry elements changes between commuting and anticommuting $\eta_\mathcal{T} \to -\eta_\mathcal{T}, \eta_\mathcal{P} \to - \eta_\mathcal{P}$. We use this freedom to require that ${\cal S}^2 = 1$ for a unitary crystalline symmetry ${\cal S}$ and to write a unitary antisymmetry, antiunitary symmetry, and antiunitary antisymmetry as the products ${\cal C} {\cal S}'$, ${\cal T}^\pm {\cal S}'$, and ${\cal P}^\pm{\cal S}'$, with ${\cal S}'$ a unitary symmetry and ${\cal T}$ and ${\cal P}$ antiunitary operators, where the superscript indicates the square of the antiunitary operator $({\cal T}^\pm {\cal S}')^2 = \pm 1$, $({\cal P}^\pm {\cal S}')^2 = \pm 1$. The combination of the tenfold-way class and the crystalline symmetry ${\cal S}$ or ${\cal T}^\pm {\cal S}'$ or crystalline antisymmetry ${\cal C}{\cal S}'$ or ${\cal P}^\pm {\cal S}'$ then uniquely specifies the crystalline symmetry class. This is the notation used in Tables \ref{tab:classification} and \ref{tab:class_Mxy}--\ref{tab:class_I} and throughout the text.

\section{Topological invariants}
\label{app:f}

The protected gapless points or regions are related to topological phase transitions of the two-dimensional Bloch Hamiltonian $H_{k_z}(k_x,k_y)$ as a function of $k_z$. This topological phase transition is associated with a change of a topological invariant. Below, we give explicit expressions for the topological invariants associated with the symmetry classes featured in this article. Since the topological phase transitions take place at $0 < |k_z| < \pi$, it is sufficient to consider the crystalline symmetry classes listed in the third column of Tabs.\ \ref{tab:class_Mxy}--\ref{tab:class_I}. The corresponding crystalline symmetry class of the full three-dimensional gapless phase is given in the first column of these tables.

\subsection{With mirror symmetry $\mathcal{M}^{x/y}$}

For definiteness, we consider the mirror operation ${\cal M}^x$.

\emph{Class $\text{AIII}^{\mathcal{M}_{+}}$}.--- At the mirror-symmetric lines $k_x = 0$ and $k_x = \pi$, the two-dimensional Hamiltonian $H_{k_z}(k_x,k_y)$ can be block-diagonalized into blocks $H_{k_z}^{\pm}(k_x,k_y)$ with even and odd mirror parity, such that each block is in class AIII. From the four one-dimensional Hamiltonians $H_{k_z}^{\pm}(0,k_y)$ and $H_{k_z}^{\pm}(\pi,k_y)$ four winding numbers \cite{chiu2016} $\nu_{k_x}^{\pm}$ can be defined, with $k_x  = 0$, $\pi$, where the superscript $\pm$ refers to the mirror parity.
A weak first-order topological phase corresponding to a stack of one-dimensional topological class AIII chains with mirror parity $\pm 1$ in the $x$-direction has $\nu_{0}^\pm = \nu_\pi^\pm$.
If $H_{k_z}(k_x,k_y)$ is gapped and not in one of the aforementioned weak topological phases, these winding numbers satisfy the constraint $\nu_{0}^+ + \nu_{0}^{-} = \nu_{\pi}^+ + \nu_{\pi}^- = 0$.
The topological invariant $\nu \in \ZZ$ describing transitions between the second-order phases and between the second-order phase and the trivial phase then reads
\begin{equation}
  \nu = \frac{1}{2}\left( \nu_{0}^+ - \nu_{0}^{-} - \nu_{\pi}^+ + \nu_{\pi}^-\right).
\end{equation}

\emph{Class $\text{A}^{\mathcal{CM}}$}.--- For this class, the lines $k_{x}=0$ and $k_x =\pi$ are in Cartan class AIII, so that one may define a Chern number \cite{chiu2016} $\nu_{\rm C} \in \ZZ$ on the full Brillouin zone as well as winding number $\nu_{k_z} \in \ZZ$ for the one-dimensional Hamiltonians $H_{k_z}(0,k_y)$ and $H_{k_z}(\pi,k_y)$. These satisfy the constraint $\nu_{\rm C} = \nu_{0} - \nu_{\pi} \, \mod 2$. For transitions not involving a first-order phase, we therefore have $\nu_{0} = \nu_{\pi}\, \mod 2$. The topological invariant $\nu \in \ZZ$ describing transitions between the second-order phases and between the second-order phase and the trivial phase then reads
\begin{equation}
  \nu = \frac{1}{2}(\nu_{0} - \nu_{\pi}).
\end{equation}

\subsection{With rotation symmetry $\mathcal{R}^{x/y}$}

For definiteness, we consider the rotation symmetry $\mathcal{R}^{x}$. 

\emph{Class $\text{AIII}^{\mathcal{P}^{+}\mathcal{M}_{+}}$}.--- We choose the representations $U_{\cal C} = \tau_3$, $U_{{\cal P} \mathcal{M}} = \tau_3$, $U_{{\cal T} \mathcal{M}} = U_{{\cal C} {\cal P} \mathcal{M}} = \tau_0$, so that $H_{k_z}(k_x,k_y) = -\tau_3 H_{k_z}(k_x,k_y) \tau_3 = H_{k_z}(k_x,-k_y)^*$. Then, at the high-symmetry lines $k_y = 0$, $\pi$, $\ZZ_2$ first Stiefel-Whitney numbers \cite{fang2015nodal} $\nu_{0}$, $\nu_{\pi}$ may be defined from $H_{k_z}(k_x,0)$ and $H_{k_z}(k_x,\pi)$ for $-\pi \le k_x \le \pi$. The difference
\begin{equation}
  \nu = \nu_{0} - \nu_{\pi}\, \mod 2
\end{equation}
is the topological invariant that describes transitions between the second-order phase and the trivial phase.

\emph{Class $\text{A}^{\mathcal{P}^{+}\mathcal{M}}$}. A topological invariant can be defined from a generalization of the Moore-Balents argument \cite{MoorePRB2007, shiozaki2014}. We choose the representation $U_{{\cal P} {\cal M}} = \tau_0$, so that $H_{k_z}(k_x,k_y) = -H_{k_z}(k_x,-k_y)^*$. At the high symmetry lines $k_y = 0$, $\pi$, $H_{k_z}(k_x,k_y)$ is antisymmetric. To construct a topological invariant, we smoothly deform $H_{k_z}(k_x,k_y)$ such that $H_{k_z}(k_x,k_y)$ is constant for $k_y = 0$ and $k_y = \pi$. 
Such a deformation is always possible, since the fundamental group $\pi_1$ of gapped antisymmetric hermitian matrices is trivial.
After this deformation, a Chern number $\nu$ can be computed by considering $H_{k_z}(k_x,k_y)$ for $-\pi \le k_x \le \pi$ and $0 \le k_y \le \pi$. The parity of $\nu$ does not depend on the deformations used to achieve the condition that $H_{k_z}(k_x,k_y)$ be constant for $k_y = 0$ and is the sought topological invariant.

\subsection{With rotation symmetry $\mathcal{R}^{z}$}

\emph{Class $\text{A}^{\mathcal{R}}$}.--- At the four high-symmetry points $(k_x,k_y) = (0,0)$, $(\pi,0)$, $(0,\pi)$, and $(\pi,\pi)$ the bands have well-defined parity under ${\cal R}^z$. The four integer invariants $\nu_{(k_x,k_y)}$ denote the number of occupied bands of odd parity of $H_{k_z}(k_x,k_y)$ at the four high-symmetry momenta $(k_x,k_y)$. In addition, one may define the Chern number $\nu_{\rm C}$ on the full Brillouin zone. The Chern number and the integers $\nu_{(k_x,k_y)}$ satisfy the constraint $\nu_{\rm C} = \nu_{(0,0)} - \nu_{(\pi,0)} - \nu_{(0,\pi)} + \nu_{(\pi,\pi)}\, \mod 2$ \cite{hughes2011}. Hence, for transitions not involving a first-order phase, we therefore have $\nu_{(0,0)} - \nu_{(\pi,0)} - \nu_{(0,\pi)} + \nu_{(\pi,\pi)} = 0\, \mod 2$. The topological invariant $\nu \in \ZZ$ describing transitions between the obstructed atomic-limit phases and between an obstructed atomic-limit phase and the trivial phase then reads
\begin{equation}
  \nu = \frac{1}{2}(\nu_{(0,0)} - \nu_{(\pi,0)} - \nu_{(0,\pi)} + \nu_{(\pi,\pi)}).
\end{equation}

\emph{Class $\text{AIII}^{\mathcal{R}_{-}}$}.--- At the four high-symmetry points $(k_x,k_y) = (0,0)$, $(\pi,0)$, $(0,\pi)$, and $(\pi,\pi)$ the bands have well-defined parity under ${\cal R}^z$. The chiral operator ${\cal C}$ interchanges bands of opposite parity. The four integer invariants $\nu_{(k_x,k_y)}$ denote the number of occupied bands of odd parity of $H_{k_z}(k_x,k_y)$ at the four high-symmetry momenta $(k_x,k_y) = (0,0)$, $(\pi,0)$, $(0,\pi)$, and $(\pi,\pi)$. The topological invariant $\nu \in \ZZ$ describing transitions between the obstructed atomic-limit phases and between an obstructed atomic-limit phase and the trivial phase then reads
\begin{equation}
  \nu = \nu_{(0,0)} - \nu_{(\pi,0)} - \nu_{(0,\pi)} + \nu_{(\pi,\pi)}.
\end{equation}
(Note that this class also allows for the definition of winding numbers for $H_{k_z}(k_x,k_y)$ along cuts through the two-dimensional Brillouin zone at constant $k_x$ or $k_y$. Nontrivial winding numbers signal weak phases.)

\subsection{With inversion symmetry $\mathcal{I}$}

\emph{Class $\text{A}^{\mathcal{T}^{+}\mathcal{R}}$}.--- Without loss of generality, we may choose the representation $U_{{\cal T}{\cal R}} = \tau_0$, so that $H_{k_z}(k_x,k_y) = H_{k_z}(k_x,k_y)^*$. For real Hamiltonians, one may define a topological invariant $\nu \in \ZZ_2$ as the second Stiefel-Whitney number \cite{fang2015nodal, ahnPRL2018}. This is the sought topological invariant.

\emph{Class $\text{AIII}^{\mathcal{T}^{+}\mathcal{R}_{-}}$}.--- Without loss of generality, we may choose the representations $U_{\cal C} = \tau_2$, $U_{{\cal T}{\cal R}}=\tau_0$, so that $H_{k_z}(k_x,k_y) = H_{k_z}(k_x,k_y)^* = -\tau_2 H_{k_z}(k_x,k_y) \tau_2$. Again, one may define a topological invariant $\nu \in \ZZ_2$ as the second Stiefel-Whitney number \cite{fang2015nodal, ahnPRL2018}. This is the sought topological invariant.

\bibliographystyle{apsrev4-2}
\bibliography{refs_HOTS}

\end{document}